%% file: Main.tex
\documentclass[acmsmall,screen]{acmart}
\usepackage{booktabs} 
\usepackage{caption}
\usepackage{adjustbox}
\usepackage{subcaption}
\usepackage{relsize}
\usepackage{float}
\usepackage{url}
\usepackage{epstopdf}
\usepackage{graphics,graphicx}
\usepackage{balance}
\usepackage{array}
\usepackage{pifont}
\usepackage{multirow}
\usepackage{xcolor}
\usepackage{amsmath}
\usepackage{footmisc}
\usepackage[normalem]{ulem}
\usepackage{paralist, tabularx}

\newcommand{\SpSeller}[0]{\textit{Related Sellers}}
\newcommand{\buybox}[0]{Buy Box}
\newcommand{\new}[1]{\textcolor{black}{#1}}

\newcommand{\red}[1]{\textcolor{red}{#1}}

\newcommand{\final}[1]{\textcolor{black}{#1}}

\newcommand{\cmark}{\textcolor{blue}{\ding{51}}}%
\newcommand{\xmark}{\textcolor{red}{\ding{55}}}

\renewcommand\footnotetextcopyrightpermission[1]{} 
\makeatletter
\def\@copyrightspace{\relax}
\makeatother

\begin{document}

\title{Investigating Nudges toward Related Sellers on E-commerce Marketplaces: A Case Study on Amazon}

\author{Abhisek Dash}
	\affiliation{
		\institution{Max Planck Institute for Software Systems}
		\country{Germany}
	}
	
	\author{Abhijnan Chakraborty}
	\affiliation{
		\institution{Indian Institute of Technology Kharagpur}
		\country{India}
	}

	\author{Saptarshi Ghosh}
	\affiliation{
		\institution{Indian Institute of Technology Kharagpur}
		\country{India}
	}
	
	\author{Animesh Mukherjee}
	\affiliation{
		\institution{Indian Institute of Technology Kharagpur}
		\country{India}
	}
	
	\author{Krishna P. Gummadi}
	\affiliation{
		\institution{Max Planck Institute for Software Systems}
		\country{Germany}
	}

\renewcommand{\shortauthors}{Abhisek Dash et al.}

\begin{abstract}
E-commerce marketplaces provide business opportunities to millions of sellers worldwide. Some of these sellers have special relationships with the marketplace by virtue of using their subsidiary services (e.g., fulfillment and/or shipping services provided by the marketplace) -- we refer to such sellers collectively as \SpSeller{}. When multiple sellers offer to sell the same product, the marketplace helps a customer in selecting an offer (by a seller) through (a)~a default offer selection algorithm, (b)~showing features about each of the offers and the corresponding sellers (price, seller performance metrics, seller's number of ratings etc.), and (c)~finally evaluating the sellers along these features. In this paper, we perform an end-to-end investigation into how the above apparatus can nudge customers toward the \SpSeller{} on Amazon's four different marketplaces in India, USA, Germany and France. We find that given explicit choices, customers' preferred offers and algorithmically selected offers can be significantly different. We highlight that Amazon is adopting different performance metric evaluation policies for different sellers, potentially benefiting \SpSeller{}. For instance, such policies result in notable discrepancy between the actual performance metric and the presented performance metric of \SpSeller{}. 
We further observe that among the seller-centric features visible to customers, sellers' number of ratings influences their decisions the most, yet it may not reflect the true quality of service by the seller, rather reflecting the scale at which the seller operates, thereby implicitly steering customers toward larger \SpSeller{}. Moreover, when customers are shown the rectified metrics for the different sellers, their preference toward \SpSeller{} is almost halved.
We believe our findings will inform and encourage further deliberation toward more effective governance of such design choices and policies adopted by e-commerce marketplaces.
\end{abstract}

\maketitle
\input{introduction.tex}

\input{related.tex}
\input{fairness-concerns.tex}

\input{Methodology.tex}
\input{buybox.tex}
\input{surveyAllCountries.tex}
\input{strike.tex}

\input{interpret-Policy-New.tex}
\input{sellerMetric.tex}
\input{discussion.tex}

\begin{acks}
	This research is supported in part by a European Research Council (ERC) Advanced Grant for the project ``Foundations for Fair Social Computing", funded under the European Union's Horizon 2020 Framework Programme (grant agreement no. 789373), and by a grant from the Max Planck Society through a Max Planck Partner Group at IIT Kharagpur. 
	\end{acks}

\bibliographystyle{ACM-Reference-Format}
\bibliography{Main}
\input{supplementary.tex}

\end{document}

%% file: introduction.tex
\section{Introduction}
\label{Sec:Intro}
E-commerce marketplaces like Amazon, Walmart and Alibaba have emerged as essential online platforms providing livelihood to millions of sellers and producers~\cite{Kaziukenas2019Amazon}.
Some of these sellers may have special relationships with the marketplace organization for using their subsidiary services~\cite{khan2016amazon, Malik2021Alpha, Dalal2016Flipkart}. 
For example, sellers can be of three broad types in Amazon marketplace: (a)~\textbf{Amazon fulfilled}--orders are fulfilled by Amazon; (b)~\textbf{Amazon shipped}--orders are only delivered by Amazon; (c)~\textbf{Merchant fulfilled}--orders are fulfilled by the seller and delivered by other third party delivery service provider (e.g., FedEx, DHL etc.). 
Moreover, Amazon itself can sell products on its marketplace. In countries, where it is not legally allowed, it sells products via \textbf{Special Merchants (SMs)} where Amazon has equity stakes~\cite{khan2016amazon,Kalra2021Amazon}. 
For instance, there are two SMs in Amazon India marketplace ({\tt Amazon.in}) -- \textit{Cloudtail India} and \textit{Appario Retail Private Ltd}. 
Collectively, we refer to all these sellers (Amazon fulfilled, Amazon shipped, Amazon and its Special Merchants) having special relationships with Amazon marketplace as \textbf{\SpSeller{}}. Table~\ref{Tab: SpecialRelationships} summarizes the different kind of sellers on Amazon marketplace and their special relationship with Amazon.

The inception of these special relationships and the monetary incentive it provides for marketplace platforms have triggered policymakers to look into the anti-competitive prowess of algorithms and policies deployed by e-commerce marketplaces~\cite{Amazon2020Questions,Amazon2019Online,EU2020Antitrust}. \new{Potential preferential treatment toward sellers using Amazon's subsidiary services has been at the core of some of these investigations. 
For instance, in the Antitrust subcommittee hearings of 2019, Amazon was asked if it takes into account which sellers use its fulfillment services in its decision making. 
However, Amazon denied that its algorithms consider such factors in their decision making~\cite{Amazon2019Online}. 
In contrast, the Federal Trade Commission (FTC) of the USA has filed a lawsuit against Amazon in September 2023, where one of the primary arguments is centered around the nuance of preferential treatment of sellers using Amazon's fulfillment services~\cite{FTC2023AmazonSue}. 
While such allegations and claims continue, to the best of our knowledge, there has been no formal end-to-end study or systematic audit of such practices encompassing the different stakeholders on Amazon or any other e-commerce marketplace. Note that these algorithms and design choices are proprietary and their effects are subtle. 
Hence, understanding such systems and their effects might require involved treatment of the research questions by breaking them down to different granularities. 
This paper attempts to bridge the mentioned gap by performing a third-party audit of the different choice architectures deployed on Amazon for potential preferential treatment (if any) toward \SpSeller{}.   
}

\begin{table}[!t]
	\noindent
	\small
	\centering
	\begin{tabular}{|p{3 cm}|p{10 cm}|}
		\hline 
		\textbf{Seller category} & \textbf{Their relationship with Amazon} \\
		\hline
            \textbf{Special Merchants} & Amazon, the seller in the USA, Germany and France marketplaces. Sellers where Amazon has equity stakes in India, e.g., Cloudtail India and Appario Retail Pvt. Ltd. They use Amazon's fulfillment service for handling their logistics\\
		\hline
		\textbf{Amazon fulfilled} & Sellers who use Amazon's fulfillment service wherein Amazon picks, packs, and ships orders on behalf of the seller. \\
		\hline
		\textbf{Amazon shipped} & Sellers who pick, and pack their products on their own; however they use Amazon's delivery services to ship their products. \\
		\hline \hline
		\textbf{Merchant fulfilled} & Sellers who pick, pack and ship orders on their own or with the help of any other third party service provider, e.g., DHL, FedEx etc. They do not use Amazon's subsidiary services for handling their logistics.\\
		\hline
	\end{tabular}	
	\caption{\textbf{Different categories of sellers on Amazon marketplace and the way they handle orders. Collectively, we refer to the sellers in the first three categories as \SpSeller{}.}}
	\label{Tab: SpecialRelationships}
	\vspace{-8 mm}
\end{table}

\subsection{Important choice architectures and research questions}
In Amazon, multiple sellers can offer to sell the same product. Consequently, when a customer views a product page, the platform presents a choice architecture to help in deciding which offer (or corresponding seller) to go for. `Choice architecture' broadly describes the way different choices are presented to customers in a multi-choice setup~\cite{johnson2012beyond, mota2020desiderata, thaler2008nudge}.
In Amazon, such choice architecture has three different facets (a)~default offer selection (\buybox{}) algorithm, (b)~offer listing page, (c)~evaluation of seller performance metrics. Next, we discuss each of these facets and pinpoint the fairness concerns to contextualise the present work.

\begin{figure}[t]
	\centering
	\begin{subfigure}{0.3\columnwidth}
		\includegraphics[width= 0.8\textwidth, height=4.5cm]{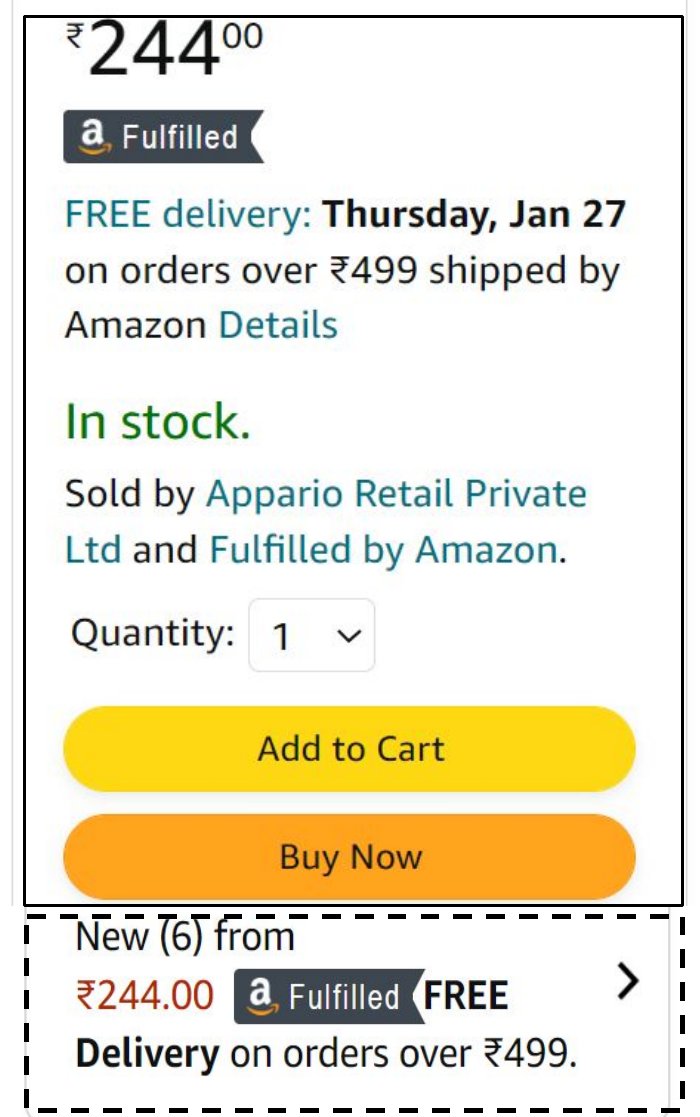}
		\vspace*{-1.5mm}
		\caption{Buy Box}
		\label{Fig: buybox}
	\end{subfigure}%
	\hfil
	~\begin{subfigure}{0.6\columnwidth}
		\centering
		\includegraphics[width= 0.8\textwidth, height=4.5cm]{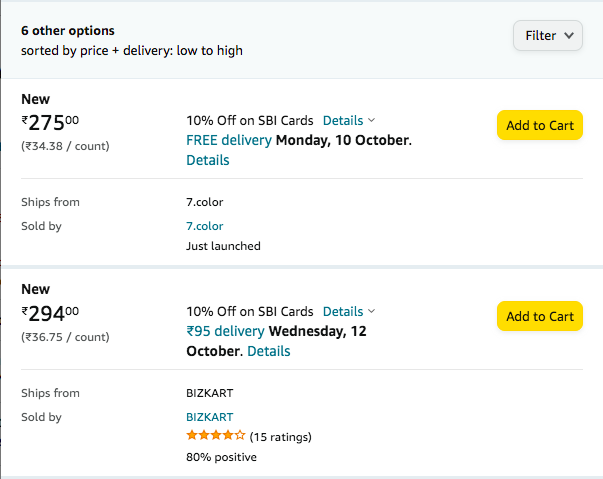}
		\vspace*{-1.5mm}
		\caption{Offer listing page}
		\label{Fig: buyboxlisting}
	\end{subfigure}
	\caption{{\bf (a)~An example of a Buy Box on Amazon. The small rectangle box (dashed), is the link to redirect customers to a separate `offer listing page'. (b)~The offer listing page, where the offers are sorted as per price + delivery charges.}}
	\label{Fig: buyboxpic}
	\vspace{-6 mm}
	
\end{figure}

\vspace{1mm} \noindent
\textbf{\buybox{} algorithm: }On each Amazon product page, one can observe a rectangular box containing two buttons -- `add to cart' and `buy now' (see Figure~\ref{Fig: buyboxpic}(a)). This box is popularly referred to as \buybox{}. When multiple sellers offer the same product, a proprietary algorithm (Amazon \buybox{} algorithm) determines which seller should be selected as a default seller i.e., the winner of the \buybox{}. 
Given the prime positioning of the \buybox{} (and the prominently displayed 
buttons inside the box), it accounts for more than 80\% sales  
on Amazon~\cite{Lanxner2021Amazon}.  
Thus, {\it winning the \buybox{} or becoming the featured offer for a product directly impacts the revenue of the sellers}. 
\new{Unsurprisingly, preferential treatment in \buybox{} offer selection has been a matter of concern among policymakers across the globe~\cite{FTC2023AmazonSue, EU2020Antitrust, House2020Judiciary, Amazon2019Online, Amazon2020Questions}. However, Amazon has maintained that its predictions reflect the customer preference~\cite{Amazon2019Online}. This leads to the first set of research questions (RQs) studied in this paper:}

\color{black}
\noindent
\textit{RQ1 (a):}~How frequently does Amazon (or a related seller) win the \buybox{} when it competes?\\
\textit{RQ1 (b):}~Do Amazon's algorithmic choices match the customers' choices when they get to decide an offer among the competing offers?
\color{black} 

\vspace{1mm} \noindent
\textbf{Offer listing page: } For every product 
with multiple interested sellers, Amazon presents a separate offer listing page 
with all offers listed for the customers to compare (see Figure~\ref{Fig: buyboxpic}(b)). For each seller, apart from the price and delivery details, 
different performance metrics such as their average user rating, \% of positive feedback, and number of ratings received (\#Ratings) are also displayed. 
These metrics are meant to convey the customer the quality of service a seller provides. 
However, \#Ratings does not necessarily indicate the true quality; rather it reflects the scale at which the corresponding seller operates. A seller that extensively sells products in multiple categories but has lower quality of service, may well have significantly higher \#Ratings as compared to a niche seller providing good quality service operating in less diverse product categories. Thus showcasing such an un-normalised attribute may steer the customer preference toward \textbf{larger} sellers (majority of whom are \SpSeller{}). This deliberation brings us to our second research question:

\noindent
\new{\textit{RQ2:} Which information on the offer listing pages influence customers' decision-making the most?}

\vspace{1mm} \noindent
\textbf{Evaluation of performance metrics:} 
Moreover, platforms may have explicit policies which may put their \SpSeller{} in a position of strength. For example, Amazon has a policy that it can \textbf{strike-through} negative feedback received by \SpSeller{}~\cite{SellerCentral2022Can}. 
As per Amazon's policy, a negative feedback concerning delivery experience can be struck through for sellers who opt for its fulfillment or delivery services (more details in Section~\ref{Sec: StrikePolicy}).
In these cases, such negative feedback does not reflect on their performance metrics. 
However, these policies are in place only for sellers using Amazon's delivery and fulfillment services i.e., \SpSeller{}. For 
other sellers (merchant fulfilled sellers), no such policies are in place. 
Hence for the exact same issue faced by a customer, different sellers on Amazon may be treated differently based on their special relationship with Amazon. This raises concerns of disparate treatment~\cite{barocas2016big} towards the sellers and is also questionable as per the recently enforced Digital Markets Act (DMA) in the European Union~\cite{EC2022DMA}. In fact, FTC's new antitrust suite on Amazon focuses on Amazon leveraging its power to reward online sellers who uses its subsidiary services~\cite{Edgerton2023Lina}.
Given the importance of seller performance metrics in the \buybox{} algorithm~\cite{Amazon2021Becoming}, such misrepresentation may create an uneven playing field for sellers (in favor of \SpSeller{}) 
by steering customer preference toward them. 
Hence to quantify and understand the awareness and effect of this review strike-through policy, our final set of RQs are as follows:

\color{black}
\noindent
\textit{RQ3 (a):}~What are the effects of Amazon's review strike-through policies on seller performance metrics reported on Amazon?\\
\textit{RQ3 (b):}~How aware are customers about these review strike-through policies and how do they interpret them?\\
\textit{RQ3 (c):}~What are the effects of Amazon's review strike-through policy on customers' choices?
\color{black}

\color{black}
\subsection{Our contributions} 
\new{In this work, 
we design a web-scraper and collect the \buybox{} information for multiple products and sellers from Amazon's marketplaces in four different countries: {\tt Amazon.in} (India), {\tt Amazon.com} (USA), {\tt Amazon.de} (Germany) and {\tt Amazon.fr} (France). 
In total, we collect and analyse data of more than 75,000+ \buybox{} competitions across 100 most frequently searched queries on Amazon. 
Our analyses over the collected data show that Amazon wins a large fraction of the \buybox{} competitions where it competes with other sellers. To further understand if this situation truly reflects customers' preferences, we conduct a survey among 200 participants (50 from each of the countries). For a set of 60 products, we show them the offers details (from different sellers) and ask them to give us their preferred choices.
We observe that the customer choices elicited in the survey do \textit{not} match with the algorithmic choices of the \buybox{} algorithm for the surveyed products in a significant fraction of cases. During the survey, customers also expressed that different seller performance metrics shown on the offer listing pages influence their decisions.
To investigate the correctness of the reported seller performance metrics, we collect seller feedback of top-1000 most active sellers on all four marketplaces and conduct an analysis of over 4M seller feedback. 
Our analyses show that Amazon's review strike-through policy results in significant discrepancy between the reported and actual performance metrics of \SpSeller{}.
To further quantify the effect of such policies on customers' decision-making, we repeat the surveys by showing them the rectified metrics. 
In this counterfactual setting, we observe that preference toward Amazon and its \SpSeller{} further reduces among customers. 
Our major findings 
are summarized below:}

\vspace{1 mm}
\noindent
$\bullet$ \textbf{Amazon \buybox{} prefer Amazon and its special merchants in the analysed competitions:} We observe that Amazon and its SMs enjoy a \buybox{} win-rate of upwards of 80\% in our collected dataset. In particular, for the data collected for USA and India, they also win more than 25\% cases, 
\textit{despite not offering the lowest price for the said product}. 
Our conducted survey for 60 products across 200 participants consistently points at significant difference between algorithmic decision and customer preferences across countries. While customers prefer Amazon and its SMs in 31\% cases, Amazon \buybox{} algorithm prefer these sellers for similar competitions in nearly 80\% cases across countries (as per the surveyed competitions). Often respondents mention seller performance metrics such as percentage of positive feedback, \#Ratings as the most influential features.

\vspace{1mm} \noindent
$\bullet$ \textbf{Amazon's review strike-through impacts seller performance metrics: }
Amazon's strike-through policy has serious ramifications on the seller performance metrics of \SpSeller{} across countries. Especially, the performance metrics of Amazon fulfilled sellers, in our dataset, drop by nearly 20\% when we include the struck-through reviews in Amazon India. 
More worryingly, 78\% of the respondents we surveyed were not even aware of such strike-through policies of Amazon.

\vspace{1mm} \noindent
$\bullet$ \textbf{Seller performance metrics have significant impact on customers' decisions:} 
Mere removal of \#Ratings feature reduces the frequency of customers preferring Amazon SMs by nearly 40\%. Furthermore, when customers are shown the rectified metrics (which takes into account the struck through reviews), the first preference votes toward Amazon SMs get almost halved and that for Amazon fulfilled sellers reduce by 16.67\%.

\color{black}

\vspace{1 mm}

\noindent
\textbf{Organization of the paper: }\new{The rest of the paper is organised as follows. Section~\ref{Sec: RelatedWorks} discusses some of the prior works in different relevant strands of research. Section~\ref{Sec: Motivation} elaborates on the main choice architectures in Amazon marketplace and discusses the potential concerning them to further motivate the work. 
Section~\ref{Sec: Methodology} gives an overview of the methods adopted in this paper. 
Sections~\ref{Sec: buybox},~\ref{Sec: SurveyBuyBox},~\ref{Sec: StrikeThrough} and~\ref{Sec: SurveyBuyBoxWithoutRateNum} elaborate more about each of the methods adopted, along with their insights. 
Specifically, Section~\ref{Sec: buybox} investigates potential preferences towards Related Sellers in the Buy Box predictions,
Section~\ref{Sec: SurveyBuyBox} compares the Buy Box algorithmic decisions with customer preferences,
Section~\ref{Sec: StrikeThrough} studies the effects of Amazon's policy of strike-through seller feedback, and Section~\ref{Sec: SurveyBuyBoxWithoutRateNum} quantifies the impact of seller performance metrics on customer decisions.
Finally, we conclude with a discussion summarizing the contributions, limitations, recommendations along with insights for different relevant stakeholders in the e-commerce ecosystem.}

%% file: related.tex
\section{Related work}\label{Sec: RelatedWorks}
To place our work in the perspective of prior research, we present a brief overview of 
the following lines of research: (a)~nudges for influencing customer behavior; and (b)~fairness in multi-sided digital marketplaces.

\vspace{1 mm}
\noindent
\textbf{Nudges to influence customer behavior}: 
Nudges are interventions in the presentation of choices to customers that steer them toward a desirable action~\cite{thaler2008nudge}. The effectiveness of the same has been elaborated in many different contexts ranging from enabling students to eat more healthy food to selecting better future plans in terms of savings~\cite{thaler2008nudge}. Similarly, nudges have been studied to be effective in steering users' decision making~\cite{schneider2018digital} and for enabling equitable donation~\cite{mota2020desiderata}. 
At the same time, many prior works have studied how companies exploit customers’ cognitive limitations and biases for their profit~\cite{hanson1999taking}. What information is available or accessible~\cite{kahneman1982judgment} and how the information is framed~\cite{tversky1981framing, schneider2018digital} plays a significant part in their decision making process. Lately, such market manipulation has been intensified by digital marketplaces~\cite{calo2013digital}. For example, unlike physical stores of yesteryear, digital marketplaces have improved the scale and sophistication of what customer behaviour information they capture and retain and how do they use it~\cite{khan2016amazon}. At the same time, they also deploy algorithms trained on such user behaviour to mediate the interaction between different stakeholders on the marketplace~\cite{dash2021when, dash2022alexa}.
Recently studies have found some inadvertent consequences of such algorithmic mediation in terms of increasing biases~\cite{dash2021when, kulshrestha2017quantifying}, spreading misinformation~\cite{hussein2020measuring, juneja2021auditing} and hate-speech~\cite{mathew2020hate, saha2021short} and so on. 
Similarly, other studies have indicated how certain information e.g., social influence, or ratings / reviews~\cite{moser2019impulse, luo2018online, mathur2019dark} if shown on shopping websites can influence the buying behaviour of customers. To this end, the current study attempts to look into the Amazon offer listing page and how the presentation and calculation of different seller performance metrics have the ability to nudge customers toward different sellers.

\vspace{1 mm}
\noindent
\textbf{Fairness in multi-sided digital marketplaces: }
In multiple online settings the interests of both the stakeholders are of utmost importance~\cite{burke2017multisided, burke2018balanced}.
Considering the two-sided setting, many nuanced algorithms have been developed considering fairness toward both customers and producers~\cite{mehrotra2018towards,patro2020fairrec,geyik2019fairness, mehrotra2020bandit,patro2020incremental,gupta2023towards}. Most of these algorithms generate recommendations by optimizing for multiple objectives pertaining to customer satisfaction, exposure of producers, and sometimes even platform based business objectives. They try to reconcile these objectives under constraints such that several fairness conditions are satisfied.

\begin{table}[!t]
	\noindent
	\scriptsize
	\centering
	\begin{tabular}{|p{1.5 cm}|p{4.5 cm}|p{1.4 cm}|p{1.4 cm}|p{1.2 cm}|p{1.2 cm}|p{1.2 cm}|}
		\hline 
		\textbf{Paper} & \textbf{Primary goal} & \textbf{Customer} & \textbf{Seller} & \textbf{Platform's} & \textbf{Platform's} & \textbf{Effect} \\
            \textbf{} & \textbf{} & \textbf{Preference} & \textbf{Opportunity} & \textbf{Influence} & \textbf{Policy} & \textbf{} \\
		\hline
            \citet{dash2021when} & Investigate bias in sponsored product recommendation toward Amazon Private Label products. & \xmark  &  \xmark & \cmark & \xmark & \xmark\\
            \hline
            \citet{Yin2021Search} & Investigates bias in product search toward Amazon Private Label products. & \xmark  &  \xmark & \cmark & \xmark & \xmark\\ 
		\hline
            \citet{dash2022alexa} & Investigates fairness and interpretability issues of conversational voice search in e-commerce from customers and producers perspective. & \cmark  &  \cmark & \xmark & \xmark & \xmark\\
            \hline
            \citet{chen2016empirical} & Investigates usage and effect of algorithmic pricing on Amazon \buybox{} offer selection. & \xmark  &  \cmark & \xmark & \xmark & \cmark\\
		\hline
            \citet{he2022market} & Investigates solicitation and effect of fake product reviews on sales of different products on Amazon. & \xmark  &  \cmark & \xmark & \cmark & \cmark\\
		\hline \hline
            This work & Investigates existence of preferential treatment toward related sellers on Amazon through different design and policy choices. & \cmark  &  \cmark & \cmark & \cmark & \cmark\\
            \hline
	\end{tabular}	
	\caption{\textbf{\new{Prior audit works on different algorithmic systems of the Amazon marketplace, and their fundamental differences with the current work. Our work incorporates customer preference, sellers' opportunities, effect of the platform and its policies in an end-to-end investigation of potential preferential treatment toward related sellers. The prior works focused on only a subset of these facets in different contexts.}}}
	\label{Tab: PriorWorks}
	\vspace{-8mm}
\end{table}

\vspace{1mm}
\noindent \textbf{Present study and prior works on Amazon platform:} 
\color{black}
While the prior works are relevant to our work in broader scheme of things, there have been a few prior audits on the Amazon marketplace, that are most closely related to the current work. We list these works in Table~\ref{Tab: PriorWorks}. 
Table~\ref{Tab: PriorWorks} shows the primary goals of the individual works (including the current study) in perspective of five different aspects: (a)~preference of customers, (b)~opportunity of sellers, (c)~influence of the platforms, (d)~Amazon's policy, and (e)~their effects. 
The study of~\citet{dash2021when} investigated and proposed methodologies to quantify biases toward Amazon's private label products in sponsored related item recommendations i.e., studied the influence of platform in product recommendations. 
Along the same lines, \citet{Yin2021Search} studied similar influence and biases in the search infrastructure of Amazon. 
While both these works investigated the influence of Amazon, both were in the context of Amazon private label products and \textit{not} on the sellers who use Amazon's subsidiary services.
\citet{dash2022alexa} performed a study to understand the fairness and interpretability issues in conversational e-commerce search through Amazon Alexa. 
This study encompassed customer preferences and seller opportunities. However, this study did not consider special relationships of the entities with the marketplace (i.e., Amazon).
Similarly, two prior works discussed seller opportunities and effect of different algorithmic systems or policies on Amazon. \citet{chen2016empirical} also studied Amazon's \buybox{} algorithm (similar to what is done in the present study). However, their primary goal was to understand the effects of algorithmic pricing and \textit{not} the question of potential preferential treatment toward \SpSeller{}. 
Finally, in another recent work~\citet{he2022market}, the authors studied the event of solicitation of fake product reviews on Amazon and its effects on sales of the different products. Although our work also considers feedback reviews on Amazon, we consider \textit{seller reviews} as opposed to product reviews in the prior work~\cite{he2022market}. 

Thus, while each of the above-mentioned prior works studied different aspects in the context of different algorithms deployed on Amazon, none of them touched upon all the aspects mentioned above. 
The current work, on the other hand, tries to investigate potential preferential treatment in Amazon's choice architecture toward \SpSeller{} considering all of the mentioned aspects.
\color{black}

%% file: fairness-concerns.tex
\section{Amazon's Choice architecture and related (Un)Fairness concerns}\label{Sec: Motivation}
Next, we describe the necessary background about the Amazon platform and its choice architectures.

\subsection{Sellers on Amazon marketplace}
Amazon has nearly 3 million active sellers across the globe~\cite{Kaziukenas2019Amazon}. As mentioned in 
the introduction, these sellers can be categorized depending on how they handle their delivery logistics. 
Table~\ref{Tab: SpecialRelationships} summarizes the different kind of sellers active on Amazon marketplaces. Notice that Amazon itself also sells products on its own marketplace in many countries. However, in India, owing to Foreign Direct Investment (FDI) policy of the government, Amazon is not allowed to sell products on Amazon India. Thus, they created joint ventures -- Cloudtail India and Appario Retail Private Ltd\footnote{While Cloudtail ceased to offer products on Amazon in 2022, Appario remains in the business of selling products on Amazon. Both companies were sellers on Amazon India while the data presented in this article was collected.} -- 
to sell products on Amazon while abiding by the rule of the land~\cite{Kalra2021Amazon, Govt2018FDI}. 
As per reports~\cite{Kalra2021Amazon}, Amazon internally refers to these two sellers as `Special Merchants'\footnote{We are including Amazon, the seller, in the same umbrella term for the purpose of this paper. Henceforth, any reference to Special Merchant (SM) would mean Amazon (the seller) in its USA, France and Germany marketplaces, and Cloudtail and Appario in its Indian marketplace.}.
Collectively, we refer to sellers who have any degree of special relationship with Amazon (i.e., Special Merchants, Amazon Fulfilled and Amazon Shipped) as \SpSeller{}. Note that the degree of special relationships vary from category to category: Special Merchants$>>$Amazon Fulfilled$>>$Amazon Shipped.

\vspace{1 mm}
\noindent
\textbf{(Un)Fairness concerns: }
Since Amazon directly (or through its special merchants in India) sells to customers, it is in direct competition with other third party sellers on its marketplaces. Similarly, Amazon also charges fees from the Amazon fulfilled sellers and Amazon shipped sellers in return of its fulfillment and delivery services~\cite{Amazon2023FBA}. Thus, higher sales for such \SpSeller{} services also provides additional monetary incentive for Amazon to prefer them through its choice architecture. Such vertical integration in the functioning of the marketplace has raised concerns among policymakers across the globe~\cite{Amazon2019Online, Amazon2020Questions, EU2020Antitrust}. 

Throughout the remainder of the section, we shall discuss the different facets of Amazon's choice architecture and describe the fairness concerns in their context.

\subsection{Amazon \buybox{}}\label{Sec: BBFconcerns}
Multiple sellers (including Amazon's \SpSeller{}) can offer to sell the same product on Amazon. These sellers compete to win the \buybox{} (or become the featured offer) for that product when a customer views the product page. In such cases, 
the \buybox{} algorithm selects which seller's offer is displayed on the \buybox{} of the product page. 
Customers typically purchase most products on Amazon 
through the \textit{\buybox{}} or \textit{featured offer} on the Amazon product pages~\cite{Lanxner2021Amazon}. (see Figure~\ref{Fig: buyboxpic}(a)). 

\vspace{1 mm}
\noindent
\textbf{(Un)Fairness concerns: }By selecting the default seller, \buybox{} provides sellers the opportunity to enhanced sales and revenue~\cite{dash2022alexa}. 
Thus, if the \buybox{} winners are not rotated properly (i.e., if the winner distribution is very skewed), only a particular seller or group of sellers will get the opportunity to sales, and a significant set of other competing sellers for the same product will be deprived of such opportunities. 
Coming from the customers' side, if a better offer at a lower price or from a seller with better quality of service is available and the \buybox{} algorithm selects a different seller by default it may lead the customer to get the same product at a higher price or from a lower quality seller and potentially lead to customer dissatisfaction.
To this end, recently, Amazon's \buybox{} algorithm has come under intense scrutiny by policymakers across the globe~\cite{EU2020Antitrust,Amazon2019Online,Amazon2020Questions}. In a recent antitrust subcommittee hearing in the USA, Amazon was asked a number of explicit questions pertaining to the \buybox{} algorithm~\cite{Amazon2020Questions}. 
Amazon stated that it refines the predictions to reflect customer preferences by considering factors such as price, fulfillment speed, seller performance etc. Amazon has also stated that it \textit{does not} consider profitability in the design of the \buybox{} algorithm~\cite{Amazon2020Questions}.

\subsection{Offer listing page on Amazon} \label{Sec: OLPFconcerns}
While the winner of the \buybox{} is shown on the product page, if a customer wants to compare all the offers for a particular product (s)he needs to click a link on the product page which reads `new ($xx$) from'. This directs to a web-page where the rest of the offers are listed in an ordered fashion.
Figure~\ref{Fig: buyboxpic}(b) shows an instance an offer listing page.
By default, the offers are shown in the increasing order of their (price + delivery).
Apart from showing the seller name, their price offer and delivery options this page also shows several seller performance metrics, e.g., average user ratings, \% of positive feedback over past 12 months, lifetime number of ratings of the sellers etc.

\vspace{1 mm}
\noindent
\textbf{(Un)Fairness concerns: }
Although some of the metrics are normalised and quantify the quality of service, features such as number of ratings (\#Ratings) are un-normalised and reflect the number of feedback ratings received by the corresponding seller in its entire lifetime on Amazon. Implicitly, \#Ratings feature (the way it is being used) is biased toward sellers who operate at larger scale, i.e., sellers who sell multitude of products across multiple categories. Hence showing such a feature may steer customers away from small-scale sellers who sell products from a specific category. 

\subsection{Seller performance metrics on Amazon}\label{Sec: StrikePolicy}

\begin{figure}[t]
	\centering
	\begin{subfigure}{\columnwidth}
		\centering
		\includegraphics[width= 0.6\textwidth]{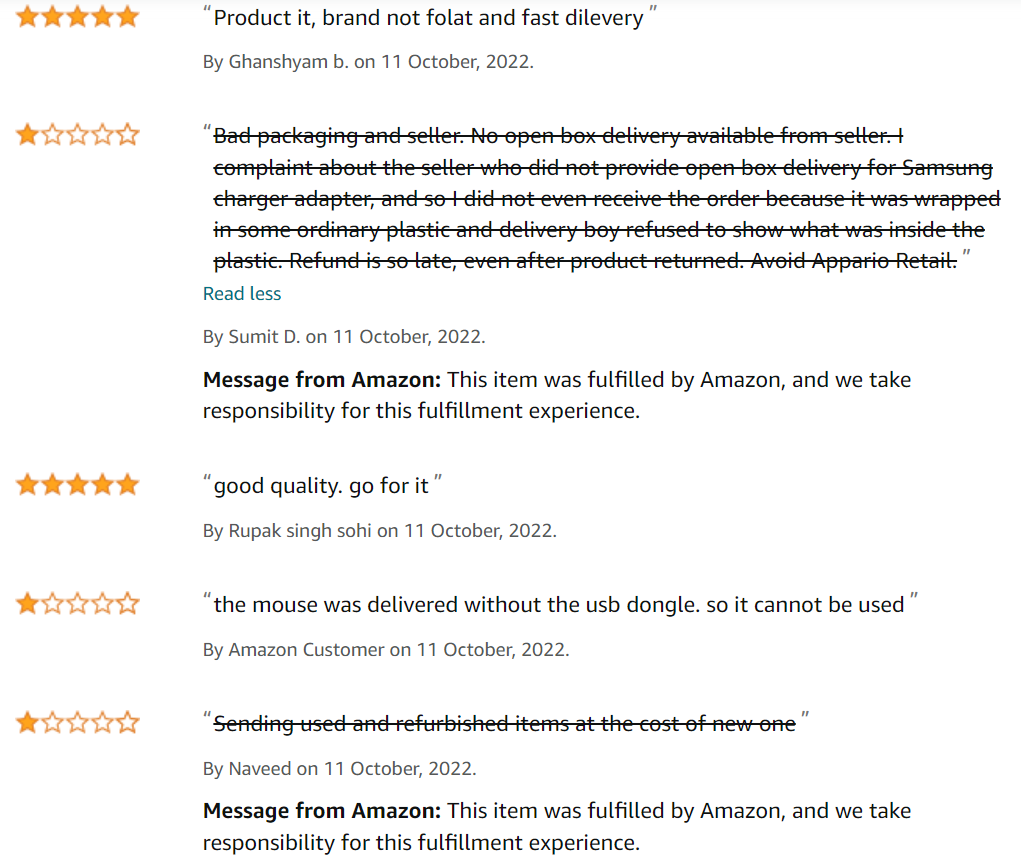}
		\vspace*{-1.5mm}
	\end{subfigure}
	\caption{{\bf Example of review striking through: the second and the last reviews have been struck through. A message from Amazon is mentioned taking responsibility of the issue. The percentage of positive feedback drops from 67\% to 40\% if we include the struck through reviews in the evaluation.}}
	\label{Fig: SellerProfilePage}
	\vspace{-6mm}
\end{figure}

The aforementioned performance metrics are evaluated from the seller feedback received by the corresponding sellers.
Each of the seller feedback comes with a rating and an associated review.
Any feedback with ratings 4 stars or above is a positive feedback; feedback with 2 stars or below is a negative feedback while feedback with 3 stars is considered to be neutral~\cite{SellerCentral2022Feedback}. Thus \% of positive feedback is the ratio of number of feedback with 4 stars or above and total number of feedback. 
Usually Amazon shows the \% of positive feedback of the past 12 months for active sellers with more than 10 buyer feedback in the same time interval. 
Average user rating is then evaluated as the mean of all the ratings received by the seller. 
Along with these normalised metrics, Amazon also shows the un-normalised lifetime number of ratings received by sellers across all categories~\cite{SellerCentral2022Feedback}.

\vspace{1mm} \noindent
\textbf{Amazon, the seller, has no associated metrics:} The interesting thing to note here is that such seller performance metrics are only available for third-party sellers on Amazon. Amazon, the seller does not have any dedicated seller profile page, nor does it have any list of feedback as shown in Figure~\ref{Fig: SellerProfilePage}. In Amazon's USA, German and France marketplaces when Amazon competes with other third party sellers, customers do not get to see its seller performance metrics.

\vspace{1 mm}
\noindent
\textbf{Amazon's feedback strike-through policy}
Amazon gives this opportunity to some Related sellers that if they request to strike-through a seller feedback, Amazon may oblige. For example, Amazon may strike-through a seller feedback if \textit{``the entire comment relates explicitly to delivery experience for an order fulfilled by Amazon (FBA)''} or \textit{``the entire comment is related to a delayed or undelivered order, which you shipped on time by using Buy Shipping''}~\cite{SellerCentral2022Can}. In both cases, in addition to the strike-through, Amazon puts up a message taking responsibility for the inconvenience faced by the customer. Furthermore, Amazon also states in its FBA advertisements that such negative reviews (which have been struck through) will \textbf{not} reflect on sellers' performance metrics~\cite{SellerCentral2022Buyer, SellerCentral2022BuyShip}. 

Figure~\ref{Fig: SellerProfilePage} shows examples of struck-through reviews where the second and the last feedback have been struck-through by Amazon. While the second feedback mentions packaging and refund issues, the last one complains about used product being delivered. 
Notice, if we disregard those two feedback completely (which is what Amazon does for related sellers), the seller has a 67\% positive feedback (2 feedback with 4 stars or above out of 3). 
However, if we consider all the reviews (i.e., now 2 out of 5 are positive), the percentage of positive feedback drops to 40\% (reduction of 27\%). 

\vspace{1 mm}
\noindent
\textbf{(Un)Fairness concerns due to special policy for \SpSeller{}:}
The strike-through policy is different for different sellers based on the special relationship that they share with Amazon and is unavailable for independent sellers 
(e.g., merchant fulfilled sellers). Hence, for the same customer inconvenience, sellers from 
different categories may be treated differently by Amazon. 
Given that a struck through review does not affect the performance metrics of the sellers, this puts Related Sellers in a beneficial position. If we disentangle Amazon -- the e-commerce platform from Amazon -- the fulfillment service provider, then, Amazon -- the platform should not put differential policies 
to ensure fair competition in the marketplace. Such policies not only violates the doctrine of \textit{disparate treatment}~\cite{barocas2016big} but also locks horn with some of the suggestions that the recently agreed upon Digital Marketing Act has (Article 6.5 of DMA)~\cite{EC2022DMA} which asks gatekeeper platforms (such as Amazon) to refrain from favorably treating products and services provided by themselves. 
It states, \textit{``The gatekeeper shall not treat more favourably, in ranking and
related indexing and crawling, services and products offered by the gatekeeper
itself than similar services or products of a third party.''}. In the USA too, the FTC plans to bring up a new antitrust suite on Amazon for leveraging its power to reward online sellers who uses Amazon's subsidiary services~\cite{Edgerton2023Lina}.

Such different policies for evaluation of seller performance metric essentially makes customers vulnerable to select sellers with lower quality of service because they are being shown the misrepresented performance metrics of sellers (see the example in Section~\ref{Sec: StrikePolicy}). Further, if such negative feedback do not reflect on the performance metric of sellers on Amazon then that affects the accountability and the quality of service to be provided to customers. This may also start a feedback loop wherein Amazon \buybox{} algorithm may potentially select \SpSeller{} who take services from Amazon -- the fulfillment service provider. In turn, Amazon -- the platform may strike-through the negative feedback such sellers may end up having; this in turn will boost their overall performance metric. Given Amazon \buybox{} algorithm also gives importance to seller performance metrics~\cite{Amazon2021Becoming}, these sellers will have higher likelihood of being selected as default sellers and so on. 

However, amid all these concerns, Amazon has consistently denied the presence of any systemic preferential treatments toward itself and its \SpSeller{}~\cite{Amazon2019Online, Amazon2020Questions, Kalra2021India2}.
{\it The lack of any systematic investigation of such design choices makes it difficult to detect evidences for such preferential treatment (if any). The methods developed in this work can be viewed as a first important step to bridge this gap.}

%% file: Methodology.tex
\begin{figure}[t]
	\centering
	\begin{subfigure}{\columnwidth}
		\centering
		\includegraphics[width= 0.6\textwidth, height=6cm]{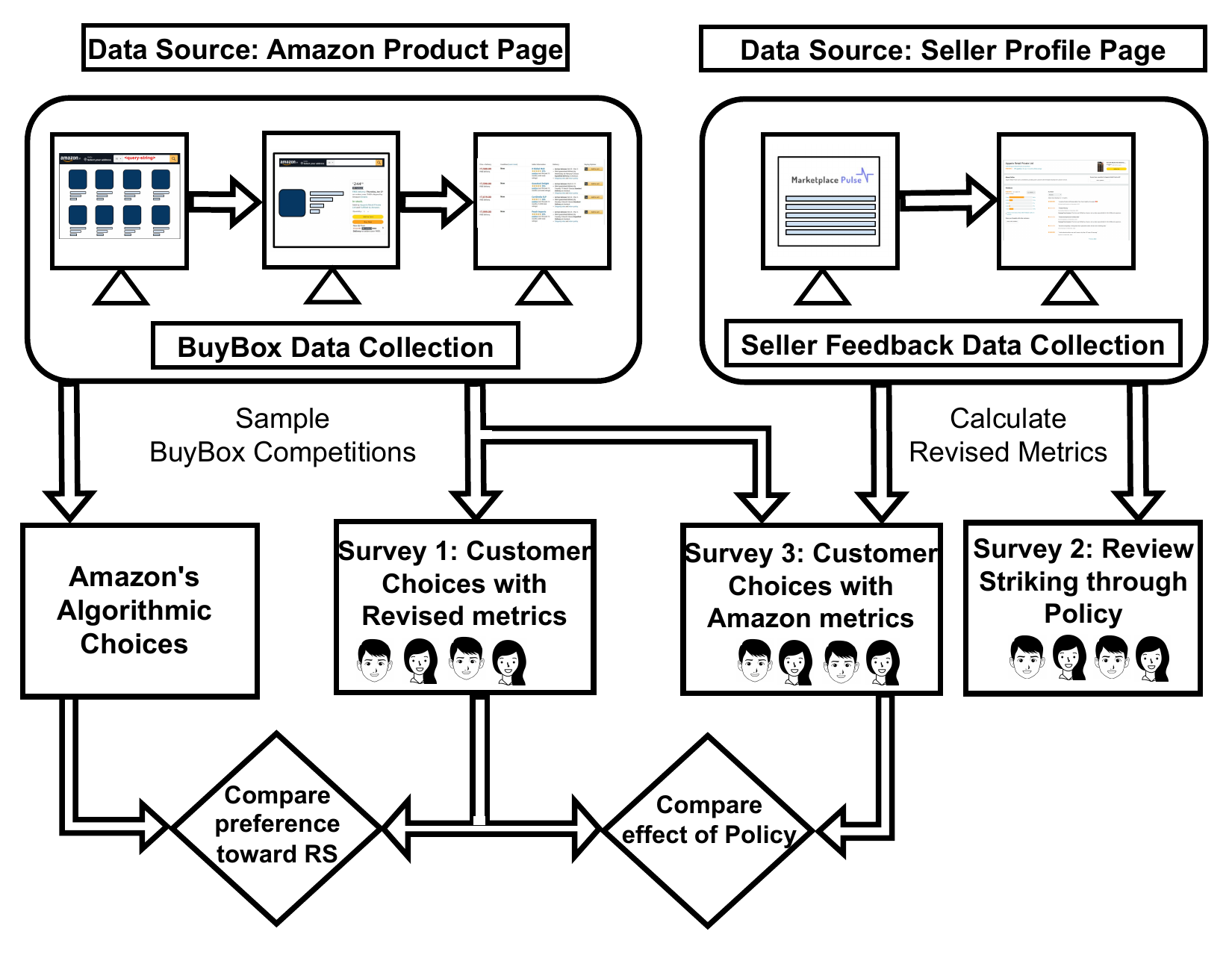}
		\vspace*{-1.5mm}
	\end{subfigure}
	\caption{{\bf Block diagram summarizing the methodology adopted in this work. RS: \SpSeller{}.}}
	\label{Fig: MethodologyPipeline}
	\vspace{-6 mm}
\end{figure}

\color{black}
\section{An overview of the research methodology} \label{Sec: Methodology}
Figure~\ref{Fig: MethodologyPipeline} shows an overview of the methodology adopted in this work to investigate for potential preferential treatment toward Amazon's \SpSeller{}. 
The primary steps in the process are -- (I)~\buybox{} data collection, (II)~survey~1: customer choices with Amazon metrics, (III)~seller feedback data collection, (IV)~survey~2: review strike-through policy, and (V)~survey~3: customer choices with rectified metrics. In the rest of this section, we elaborate each of the steps. 

\noindent
\textbf{Step I: \buybox{} data collection:} To begin our analyses, we collect data from Amazon marketplaces functioning in four different countries: India ({\tt Amazon.in}), USA ({\tt Amazon.com}), Germany ({\tt Amazon.de}), and France ({\tt Amazon.fr}). 
Our selection of these marketplaces is motivated by the recent popular antitrust investigations that Amazon faced in each of these countries and the fact that they are among the biggest and emerging marketplaces of Amazon in the respective continents~\cite{Amazon2020Questions,Amazon2019Online,EU2020Antitrust,Kalra2021Amazon,Schmitt2023Amazon}. First, we conduct a product search on an Amazon platform, then visit the product page of each search result and collect the \buybox{} competition data if multiple sellers offer to sell the same product. Since \buybox{} winners tend to change with time, we collect data for different \buybox{} competitions over a period of two weeks. In the first step, we shall look into these competitions and see how frequently does Amazon (or its SMs in India) win \buybox{} competitions when it competes for them. 

\noindent
\textbf{\final{Step II: Survey~1: Customer choices with Amazon metrics:}} 
Even if Amazon and its SMs win a larger fraction of \buybox{} competitions, one argument can be that the algorithmic decisions are reflective of the customers' preferences. In fact, this argument has been a common response of Amazon across different antitrust hearings (that their algorithmic decisions are based on customer preferences)~\cite{Amazon2019Online}. 
To verify if there is any inherent preference of customers toward Amazon and its special merchants, we conduct a customer survey over a sample of competitions drawn from our collected dataset.
We get the customer choices on which seller/offer they would prefer for these competitions. Then we compare the customer choices elicited from this survey with the frequency of times those \buybox{}es were actually won by Amazon and its special merchants in our collected dataset (over a period of two weeks). 
If the customer choices and algorithmic choices are not similar to each other, then we can conclude that Amazon's algorithmic decisions do not necessarily reflect the customer preferences. During the same survey, we also ask the respondents regarding which features of the different offers (i.e., their sellers) influences their decision the most.

\noindent
\textbf{Step III: Seller feedback data collection:} Next, we check how the Amazon policies affect the sellers. 
Seller performance metrics are important in Amazon \buybox{} algorithmic decision-making~\cite{Amazon2021Becoming}. Similarly, given the amount of seller performance metrics shown on offer listing pages, these metrics can affect customers' decision making as well. However, Amazon's review strike-through policy can play a pivotal role in this context, as it is different for different type of sellers based on how they handle their delivery logistics. 
To understand how review strike-through can affect the seller performance metrics, we collect the seller feedback garnered by 1,000 active sellers on all the four Amazon marketplaces stated above. 
We evaluate different seller performance metrics for the corresponding sellers from their seller feedback data, and compare it with their Amazon reported metrics for the past one year to quantify the effect of Amazon's policies.

\noindent
\textbf{Step IV: Survey 2: Amazon's strike-through policy:}
While the method described in Step~III brings out the effect of the review strike-through policy on the metrics, it does not shed light on the knowledge of customers (who use these metrics for their decision making) about such policies. 
To this end, we conduct another customer survey among respondents to understand their knowledge and interpretation of Amazon's strike-through policies. 
In this survey, we particularly evaluate Amazon's strike-through policy along three dimensions from the perspective of customers: (a)~transparency of the policy, (b)~interpretability of the policy, and (c)~fairness of the policy.

\noindent
\textbf{\final{Step V: Survey 3: Customer choices with rectified metrics:}}
Although the survey in Step~I may bring out what features influence customers' decision making according to their self-reported explanations, it is important to quantify the exact effects. 
For example, if in Step~III, a significant discrepancy is observed between Amazon's reported metrics and the re-evaluated or rectified seller performance metrics, then there can be some potential effects of such discrepancies on customer decision making as well. Therefore, we conduct surveys similar to Step~I by putting customers in a \textit{counterfactual setup}, e.g., where \#Ratings is not shown or where seller performance metrics are shown including the struck-through reviews (rectified performance metrics for brevity). 

\vspace{1mm}
\noindent 
Thus, through empirical observations from the Amazon \buybox{} competitions and customer surveys at each step, this work aims to bring out preferential treatment (if any) toward Amazon \SpSeller{} due to Amazon's choice presentation apparatus. 
In the remainder of this paper, we shall discuss each of the steps with greater details followed by the insights from each of the steps.
\color{black}

%% file: buybox.tex
\section{RQ1 (a): Preference toward \SpSeller{} in \buybox{} predictions}\label{Sec: buybox}
We start by looking at the outcomes of \buybox{} algorithm and their preference (if any) toward \SpSeller{}. In the beginning, we focus on the preference toward sellers with the strongest special relationship, i.e., \textit{Special Merchants}: which refers to Cloudtail and Appario in the context of Amazon India and Amazon, the seller in all its other marketplaces.
Our motivation stems from the
concerns raised in the US antitrust hearings where policymakers are concerned about preferential treatment of Amazon toward itself as a seller~\cite{Amazon2019Online, Amazon2020Questions}. During the hearings, Amazon was categorically asked if its \buybox{} algorithm prefers itself over other third party sellers and Amazon had denied having any preferential treatment in reply.

\subsection{Dataset gathered for \buybox{} competitions}
As a proof of concept, we collect data from Amazon marketplaces functioning in four different countries: India ({\tt Amazon.in}), USA ({\tt Amazon.com}), Germany ({\tt Amazon.de}), and France ({\tt Amazon.fr}). 
Given the popularity of product search on e-commerce platforms~\cite{sorokina2016amazon}, we proceed to collect our dataset by navigating the product pages following the links from the search results. 
To this end, we searched a set of $100$ distinct most searched keywords (queries) on Amazon~\cite{Hardwick2021Top} and collected buy box competition data for a period of two weeks across all the four Amazon marketplaces. The data collection methodology is elaborated in Appendix~\ref{Sec: datacollectionBuyBox}.
Upon searching a query on Amazon, a list of product offers are shown on multiple result pages (SERPs), with several results per page. When you click on the products, Amazon redirects customer to the product detail page. We consider the seller who features on the product details page by default to be the \buybox{} winner.
As Buy Box winners change with time, we make it a point to {\it repeatedly} visit product pages at different timestamps.
Upon reaching a product page, we collect product metadata, e.g., average user ratings, number of reviews, title, brand, the seller that won the \buybox{} during the crawl (as was shown in Figure~\ref{Fig: buyboxpic}(a)). 
If the link to the `offers listing page' exists, i.e., if multiple sellers are offering the product, we further simulate an automated click on the link and collect metadata for all offers (as shown in Figure~\ref{Fig: buyboxpic}(b)). 
For each offer on the offer listing page, we collect their price offered, delivery charges (if specified), seller performance (avg. rating, \% of positive feedback, etc.), delivery options (FBA or not), etc. 

\begin{table}[t]
	\noindent
        \footnotesize
	\centering
	\begin{tabular}{ |p{3 cm}|p{2 cm}|p{2 cm}|}
		\hline
		{\bf Country} & {\bf \# Competitions} &   {\bf \# Sellers}\\
		\hline
		India ({\tt Amazon.in}) & 20,553 & 2766 \\
		\hline
            USA ({\tt Amazon.com}) & 12,864 & 2577 \\
		\hline
		Germany ({\tt Amazon.de}) & 14,784 & 1695 \\
		\hline
            France ({\tt Amazon.fr}) & 29,237 & 2337\\
		\hline 
	\end{tabular}	
	\caption{{\bf Statistics of the dataset collected for buy box competitions across Amazon's different marketplaces. For each country, we state the number of number of competitions ($<$product, timestamp$>$ pairs), and the total number of distinct sellers competing in the different competitions. }}
	\label{Tab: QueryStat}
	\vspace{-8mm}
\end{table}

\vspace{1 mm}
\noindent
\textbf{Competition:} For many products on {\tt Amazon} (also in our dataset), more than one seller may not be present and therefore there is \textit{no} competition for the \buybox{}. For the purpose of investigating the \buybox{} algorithm, we only consider instances where there exist multiple sellers competing for the \buybox{}. We call such instances \textit{\textbf{competitions}}. 
A competition is a \textbf{$<$product, timestamp$>$} pair, i.e., we consider each instance of visiting a particular product page as a different competition.
We analyse over 70K competitions on {\tt Amazon} over a period of 2 weeks. 
Table~\ref{Tab: QueryStat} states the list countries, and for each country, the number of competitions and the number of sellers competing in them.

\subsection{How frequently do SMs win Buy Boxes?}
When we inspect the competitions where Amazon's special merchants compete in our dataset, we observe that a huge majority of the \buybox{} competitions are eventually won by the Amazon SMs. The Amazon SMs win more than \textit{80\%} of all competitions where they are involved in our dataset (Table~\ref{Tab: BuyBoxWinSM}). This observation is consistent across all the four Amazon marketplaces we analyzed. 
Moreover, we observe that 11\%, 13.22\%, 17.74\%, and 26.23\% of other sellers competing with Amazon SMs get to win at least one \buybox{} competition throughout our data collection period in the India, USA, Germany and France marketplaces respectively. In other words, as per the competitions analysed, \buybox{} winners are more distributed in Germany and France marketplaces than in the Indian and USA marketplaces.

\begin{table}[!t]
	\noindent
        \footnotesize
	\centering
	\begin{tabular}{ |p{3.5 cm}|p{2.5 cm}|p{2.5 cm}|}
		\hline
		{\bf Country} & {\bf \% \buybox{} won} & {\bf \% \buybox{} won OOP}  \\
		\hline
		India & 87.74\%  & 29.15\% \\ 
		\hline
		USA & 86.12\% & 26.39\% \\ 
		\hline
		Germany & 85.35\% & 13.91\% \\
		\hline
		France & 79.98\%  & 12.45\% \\ 
		\hline
	\end{tabular}	
	\caption{{\bf  \buybox{} win rate of Amazon and SMs in competitions across different countries and \% of out-of-position \buybox{} won in our collected data. We observe that SMs won 80\% or more competitions wherein they competed for the buy box. In India and USA, more than 25\% of the buy boxes were won by Amazon when they were not offering the product at the lowest price either. The percentage is almost halved in the German and France marketplaces.}}
	\label{Tab: BuyBoxWinSM}
	\vspace{-8mm}
\end{table}

A natural hypothesis to explain the above observation is that the Amazon SMs may be providing better offers than the other sellers, e.g., they may be selling the products at lower prices, and hence they are winning most of the Buy Boxes. 
However, we observe that in many cases, the SMs winning the \buybox{} do \textit{not} offer the best (price + delivery charges) as mentioned in the offer listing page. We consider such instances as \textit{\textbf{out-of-position (OOP)}} \buybox{} wins. Readers can refer to Appendix~\ref{Sec: OOPBuybox} for an example of out of position \buybox{} win.

\vspace{1mm}
\noindent \textbf{Amazon SMs have a large fraction of out-of-position \buybox{} wins:} We analysed what percentage of \buybox{} wins of Amazon SMs are potentially out-of-position, i.e., where SMs are winning the \buybox{} even though they are not offering the best offer. The break-up of out-of-position \buybox{} wins are shown in the last column of Table~\ref{Tab: BuyBoxWinSM}.
In our collected dataset, we observe that in India and USA, the Amazon SMs win out of position in more than 25\% of the times they win the competition. However, the percentages reduce to almost half of it when it comes to European marketplaces such as Germany and France.
Although quantitatively the percentages are within 30\%, but 
considering the very high win rate of the Amazon SMs even such percentages account for a significant number of \buybox{} competitions.
Given the \buybox{} algorithm almost always selects a default, and the effectiveness of such default selections in nudging customers toward the selected seller, it is important to think through such algorithmic decisions thoroughly. On one hand, such outcomes can be potentially unfair to the \textit{competing sellers} who are offering the product at competitive prices and depend on Amazon's algorithms for their revenue. On the other hand, it can be potentially unfair to the \textit{customers} especially if the same product was purchasable from a different seller at a better price.

\color{black}
\vspace{1mm}
\noindent {\bf Takeaways from this section:}
The primary takeaways from this section are as follows:

\noindent
$\bullet$ We observe that Amazon and its special merchants enjoy a \buybox{} win-rate of upwards of 80\% whenever they compete for the \buybox{} (in our collected dataset). 

\noindent
$\bullet$ In particular, for the data collected for USA and India, they also win more than 25\% cases, 
\textit{despite not offering the lowest price for the said product}.
\color{black}

%% file: surveyAllCountries.tex
\section{RQ1 (b): Do customers prefer Amazon SM over other sellers?}\label{Sec: SurveyBuyBox}
Even with the results presented earlier in this section, a question / counter-argument remains -- \textit{what if the customers on Amazon have an implicit preference toward Amazon SMs, and the Amazon Buy Box algorithm is replicating that preference as per the feedback that it is receiving from customers' behavior?}
In other words, the choice of Buy Box algorithm should match with (or deviate as less as possible from) the decisions of customers.
If it does, then even if the choices are in favor of SMs even though they do not offer the lowest price (or any related seller for that matter), there may not be any existence of preferential treatment in the decision making.

\subsection{Survey setup}  \label{sub:survey1-setup}

\new{For conducting these surveys, we chose 200 participants from India, USA, Germany and France (50 from each) who have a high approval rate ($\ge$ 98\%) on Prolific crowdsourcing platform (\url{https://prolific.co/}). Our motivation for a general set of respondents stems from the fact that anyone can purchase and/or potentially sell on Amazon, and hence can potentially be affected by the algorithms and design choices of the platform. Each respondent was remunerated at a rate of 9 GBP per hour which is recommended by Prolific to be a good and ethical rate~\cite{Prolific2023Payment}.
We also selected a balanced gender break up for the survey resulting in 49\% female and 51\% male participants taking the survey. During our survey, 81\% of the participants reported to be regular shoppers on Amazon.}

We randomly selected 60 products (15 from each marketplaces) where even though Amazon SMs did not offer the lowest price, yet they ended up winning the Buy Box for these products in the first iteration of the data collection.
For each such product, we showed the respondents the top--4 offers shown on the Amazon's offer listing page for that product. In case the Amazon SM that won the Buy Box is not within the top--4 offers, we showed top--3 along with the Amazon SM. 
For each of the four selected offers, we show all the features that are shown on Amazon's offer listing page -- name of the seller, price offered (including delivery charges, if any), average user rating of the seller, percentage of positive feedback of the seller, number of ratings received, and delivery option (FBA or not). 
Note that for `Amazon, the seller' in its USA, German and France marketplaces, none of the performance metrics e.g., average rating, percentage of positive feedback or number of ratings etc., are available and Amazon itself does not show the performance metrics for `Amazon, the seller'. Hence, we leave those places blank for `Amazon, the seller' in the survey as well.

\new{Given the four offers, we asked the respondents the following questions:\\ 
\textit{``Suppose you are willing to buy a `<\textit{product name}>' on Amazon and these are the offers from different sellers for the same product. Which one would you prefer to buy from?''}. \\
Further to understand which feature(s) is(are) important for the customers to make their decision, we also asked them the following: \textit{``Briefly reason about the order of your preferences.''}\\}
We pose the first question as a multiple choice grid such that each respondent could provide an ordered preference for the four offers. 
We repeated this exercise for 15 different products (where Amazon SMs won the Buy Box out-of-position) from each of Amazon's four marketplaces and the competition (set of 4 offers) 
for each product was evaluated by 50 respondents resulting in a total of $3000$ evaluations.

\vspace{1 mm}
\noindent
\textbf{Observations:} 
For a particular respondent and a particular product, we consider a first preference vote (i.e., rank 1 vote) to be the offer chosen by the respondent for the said product. In other words, had there been no algorithmic system and the customers were to select the offers themselves, a first preference vote essentially means a vote of willingness to buy from that particular seller.
Out of the 3000 evaluations,  
in only 31\% cases respondents chose the offer from Amazon SMs as the first preference. 
In contrast, if we consider the winner of the Buy Box for the 60 products surveyed at different time points that we observed during our data collection, more than 80\% of the times the Buy Box had been won by the Amazon SMs. Thus, there is a big difference between what customers prefer, and what the Amazon Buy Box algorithm selects, with respect to Amazon SMs.

\subsection{Observations segregated by countries}

While the above observation itself is striking, it is important to study the trends segregated across countries because of the different ways Amazon might operate in different countries, and the different ways customers of different countries interact with the marketplace.

\begin{figure}[t]
	\centering
	\begin{subfigure}{0.24\columnwidth}
		\includegraphics[width= \textwidth, height=3.5cm]{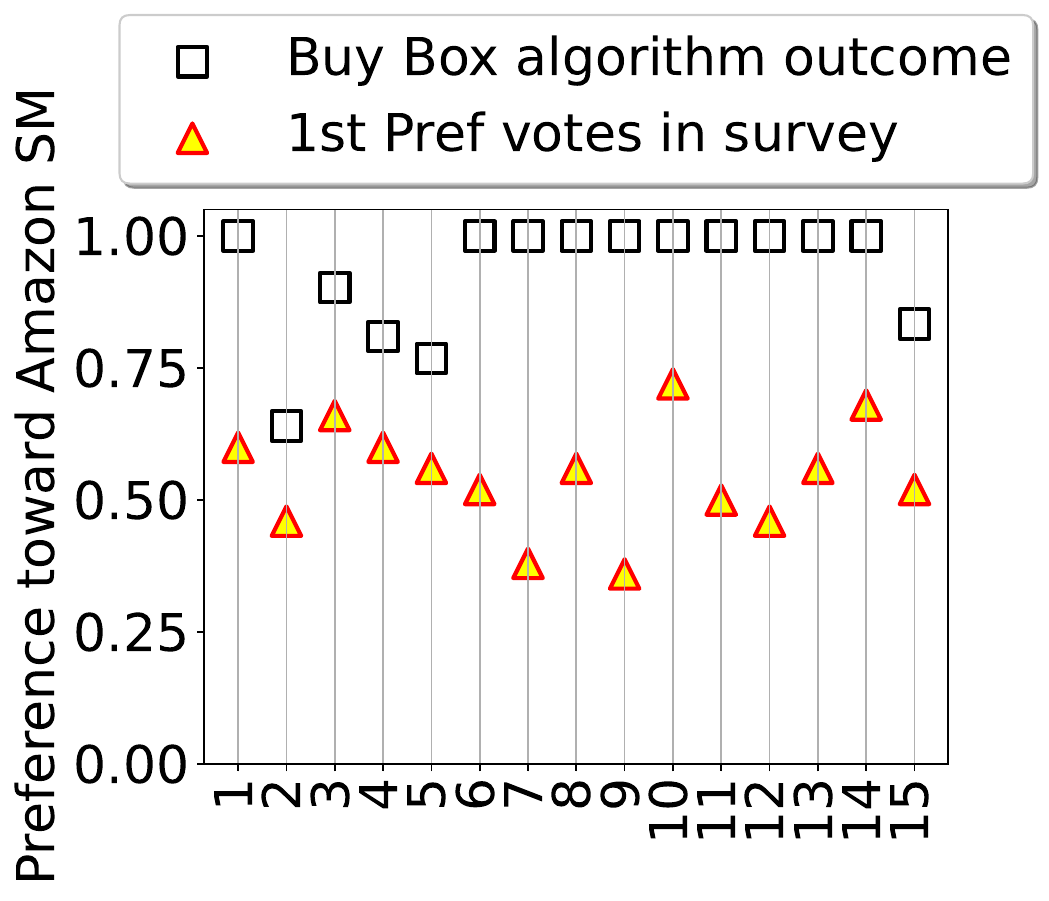}
		\vspace*{-4mm}
		\caption{India}
	\end{subfigure}%
        \begin{subfigure}{0.24\columnwidth}
		\includegraphics[width= \textwidth, height=3.5cm]{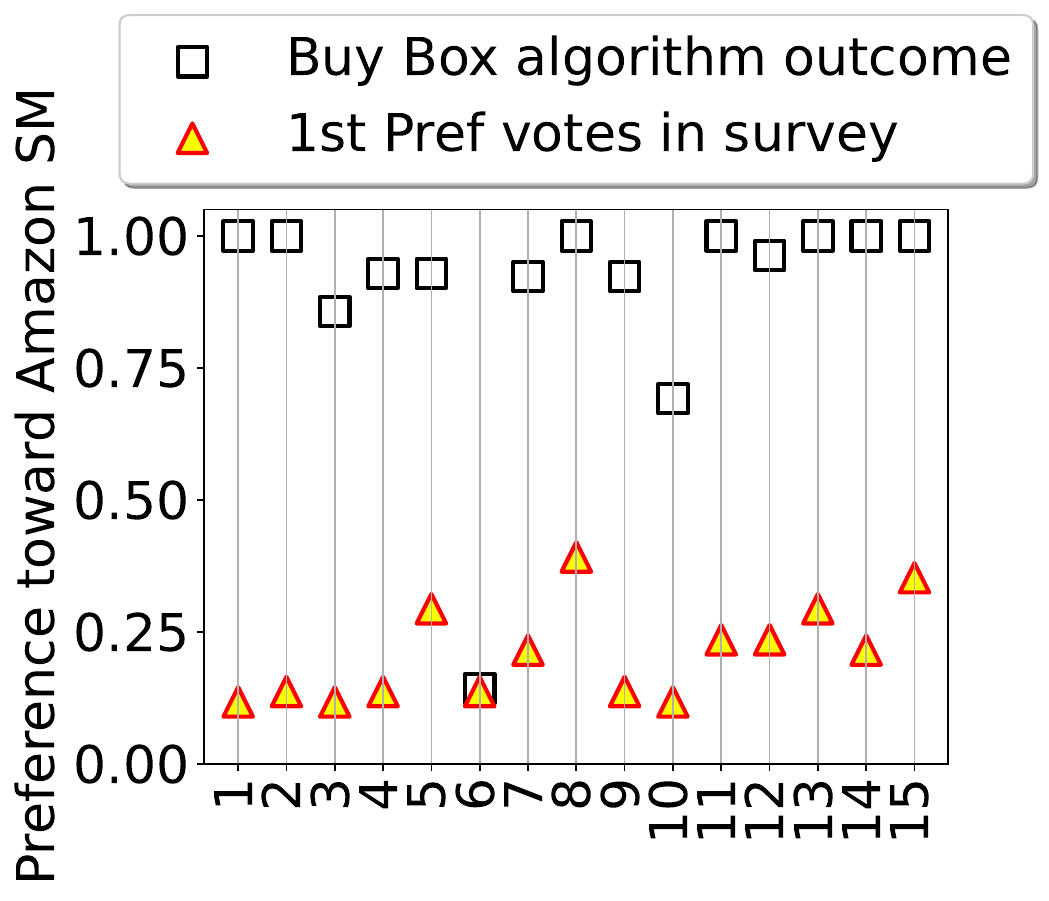}
		\vspace*{-4mm}
		\caption{USA}
	\end{subfigure}%
        \begin{subfigure}{0.24\columnwidth}
		\includegraphics[width= \textwidth, height=3.5cm]{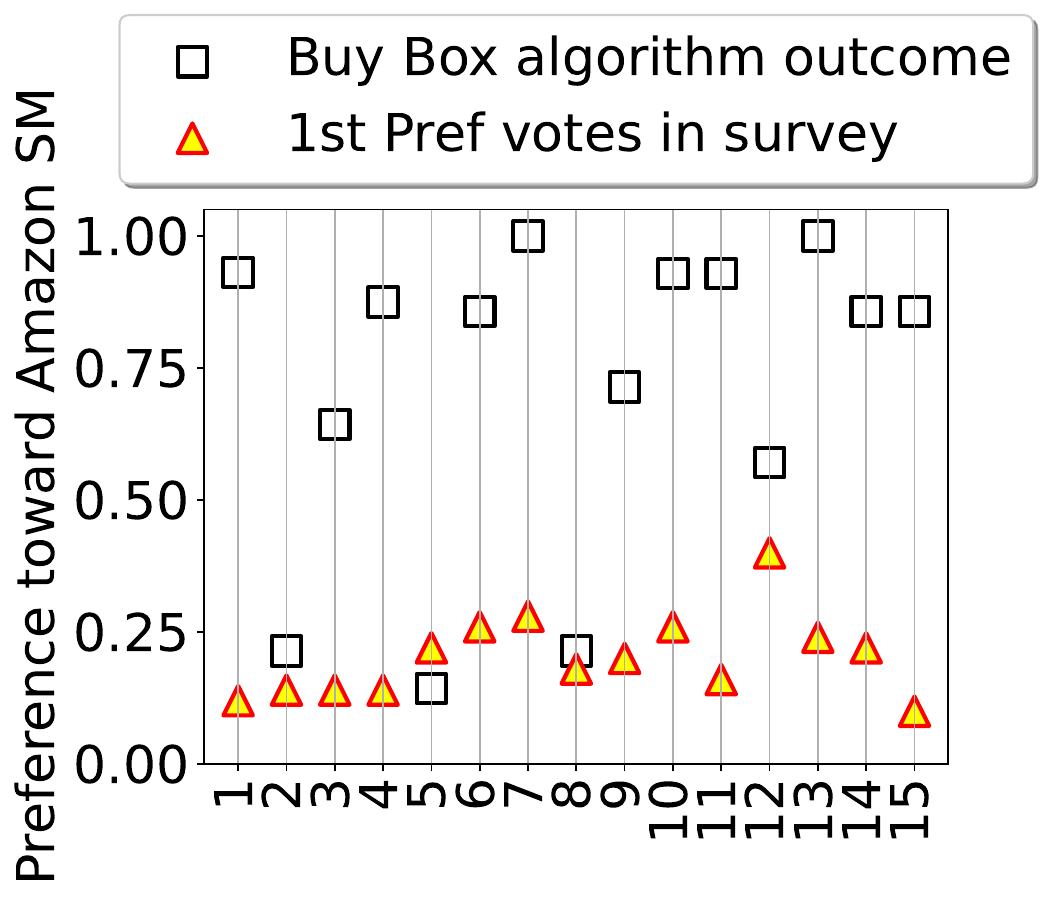}
		\vspace*{-4mm}
		\caption{Germany}
	\end{subfigure}%
        \begin{subfigure}{0.24\columnwidth}
		\includegraphics[width= \textwidth, height=3.5cm]{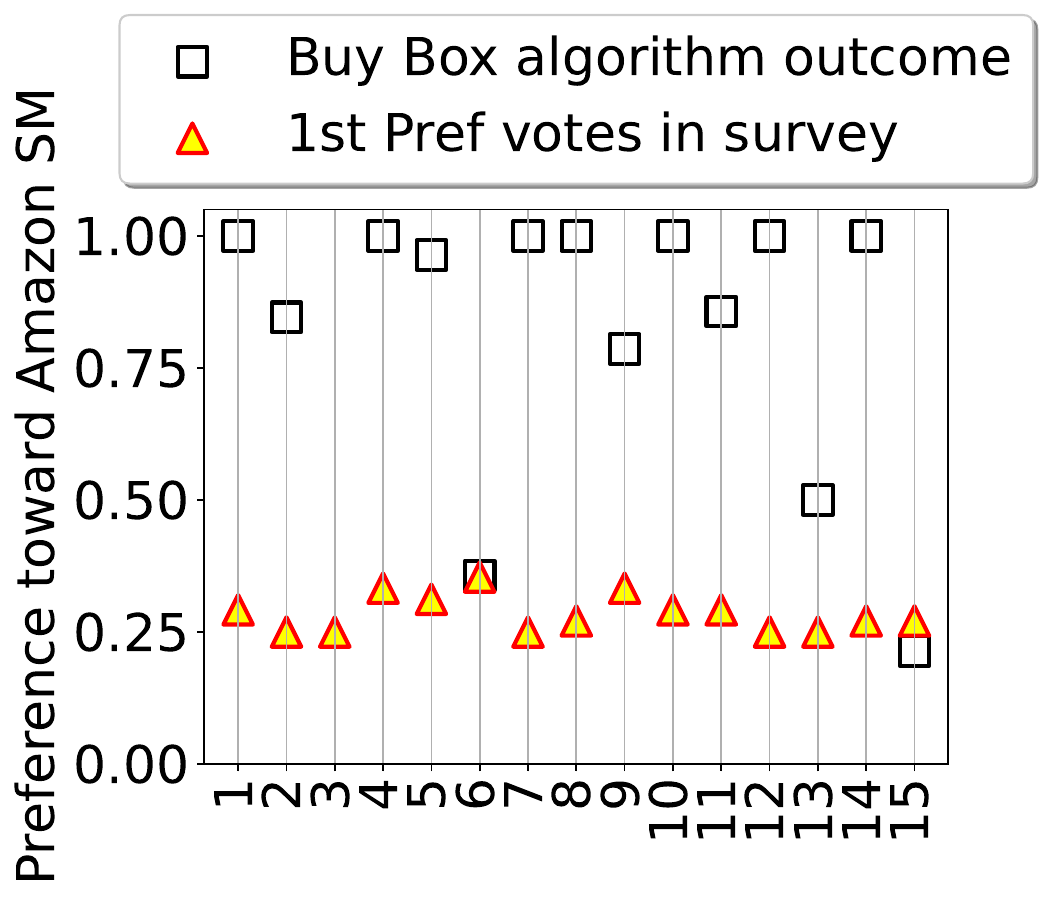}
		\vspace*{-4mm}
		\caption{France}
	\end{subfigure}%
	\vspace{-3mm}
	\caption{ {\bf Preference toward Amazon SMs for the 15 products surveyed, (i)~in Buy Box selection outcome (squares), (ii)~first preference of respondents when all features were shown (triangles) in its (a)~Indian, (b)~American, (c)~German, and (d)~France marketplaces. We consistently observe drop in the customers' preference toward SMs as opposed to the algorithm's preference.}
	}
	\label{Fig: SMPreferenceWithWithout}
    \vspace{-4 mm}
\end{figure}

\noindent
\textbf{India: }Out of the 750 evaluations done by customers of India, 407 (i.e, nearly 54\%) times the Amazon Special Merchants (i.e., Cloudtail or Appario) were given the first preference votes. However, out of all the competitions in our dataset, Amazon SMs won the \buybox{} in nearly 90\% cases. 
Figure~\ref{Fig: SMPreferenceWithWithout}(a) shows, for each of the 15 products included in the survey (along $x$-axis)
(i)~the fraction of times the Amazon SM actually won the Buy Box according to Amazon's algorithm as observed over the period of our data collection indicated by \textbf{squares},  
and 
(ii)~the fraction of respondents in our survey who voted for the Amazon SMs (first preference vote) indicated by \textbf{triangles}. 
For 10 of the 15 products, only a single seller, i.e., an Amazon SM, won the Buy Box in all competitions. However, when the competitions were decided by respondents in our survey, in none of the cases all the first preference votes went to Amazon SMs (or any single seller). 

\vspace{1mm}
\noindent
\textbf{USA: }Out of the 750 evaluations done by customers in the USA, 157 (i.e, nearly 21\%) times Amazon, the seller was given the first preference votes. However, out of all the competitions in our dataset, Amazon won the \buybox{} in 92\% cases. 
Figure~\ref{Fig: SMPreferenceWithWithout}(b) shows the observation segregated across the different products surveyed for Amazon's USA marketplace.
For 7 of the 15 products, only a single seller, i.e., Amazon, won the Buy Box in all competitions. In all the products (barring product 6), the first preference vote is significantly lower for Amazon.

\vspace{1mm}
\noindent
\textbf{Germany: }Out of the 750 evaluations done by customers from Germany, 153 (i.e, nearly 20\%) times Amazon, the seller was given the first preference votes. However, out of all the competitions in our dataset, Amazon won the \buybox{} in 74\% cases. 
Figure~\ref{Fig: SMPreferenceWithWithout}(c) shows the observation segregated across the different products surveyed for Amazon's Germany marketplace.
For only 2 of the 15 products, only a single seller, i.e., Amazon, won the Buy Box in all competitions. Notice, this corroborate with our initial observations wherein we pointed that \buybox{} winners are relatively more distributed in Germany. Apart from products 2, 5 and 6 in all the rest of the cases, we again see considerable drop in the first preference votes for Amazon as opposed to its frequency of winning the \buybox{}.

\vspace{1mm}
\noindent
\textbf{France: }Out of the 750 evaluations done by customers from France, 214 (i.e, nearly 28\%) times Amazon, the seller was given the first preference votes. However, out of all the competitions in our dataset, Amazon won the \buybox{} in 83\% cases. 
Figure~\ref{Fig: SMPreferenceWithWithout}(d) shows the observation segregated across the different products surveyed for {\tt Amazon.fr}.
For 7 of the 15 products, only a single seller, i.e., Amazon, won the Buy Box in all competitions. Apart from products 6 and 15, in all the cases here also the drop off in the two percentages (Buy Box selection vs respondent selection) are significant.

\color{black}
\subsection{RQ2: Thematic analyses of the answers mentioning the reason for preferences}

As stated in Section~\ref{sub:survey1-setup}, we also asked each of the respondents to briefly reason about their preference orders (to which they could answer using long answer texts). 
Note that user preferences are subjective in nature. Hence, to verify the truthfulness of the responses that respondents provided, minute analysis of the reasons is essential.
To this end, we went over the responses manually and figured out the following five prominent themes in the reasons provided by respondents. 

\noindent
\textbf{Seller performance metrics: } The `percentage of positive feedback' and `average user rating' of the sellers were mentioned by many respondents for their reasons to give first preference votes. Almost half of the responses from Indian participants, 30\% of the responses from the USA and Germany, and 25\% of the responses from France, mentioned these two performance metrics as important features in their consideration. 
These observations underline the importance of seller performance metrics in customers' decision making. 

\noindent
\textbf{Price: } Nearly 60\% of the respondents from USA and France mentioned price to be an influential factor in their decision making. This percentage is slightly lower for Germany (50\%) and India (40\%). This observation not only highlights the importance of price in decision making of customers, but also corroborates the findings of prior works~\cite{chen2016empirical} where it was shown that price is among the influential factors in algorithmic decision making as well.

\noindent
\textbf{Delivery option:}
Although not explicitly mentioned as many times as some of the other features,  81.2\% of the Indian respondents and 66\% of the respondents from USA stated that they opt for sellers whose offers are fulfilled by Amazon (including the Amazon SMs). Some of the responses mention FBA to be a preference, e.g., `\textit{Seller C has highest number of positive feedback and it is FBA also which is a positive}'; `\textit{Preference is based on the number of ratings and positive feedback \%. Also if it is Fulfilled by Amazon it receives a higher preference}'. 
Such preference toward Amazon FBA options is understandable since it almost always ensures a faster delivery too. However, this percentage drops to 37\% and 38\% in Germany and France respectively, indicating that customers in many countries may not necessarily have any inherent preference for sellers fulfilled by Amazon.

\noindent
\textbf{Number of ratings (\# Ratings): } Surprisingly, 60\% of the responses from Indian respondents explicitly included the mention of \#Ratings to be one of the most influential features in their decision making. Note that the two Amazon SMs in {\tt Amazon.in} sell products across all available categories and hence their \#Ratings is of the order of $10^6$; while for majority of the other sellers this is of the order of $10^3$ or $10^2$.
Thus, the mere mention of \#Ratings itself drives customers toward these sellers more. 
We did not find \#Ratings being mentioned prominently in the surveys conducted for other countries. 

\noindent
\textbf{Amazon, the seller:}
Another interesting observation in the survey of respondents from USA, Germany and France was the mention of `Amazon' in the reasoning. In 22\%, 20\% and 28\%  cases respondents explicitly mentioned Amazon in their explanation in those countries respectively. Some of the explanations are, `\textit{I selected Amazon because of their good reputation. I then selected based on best price.}'; `\textit{I chose Amazon because I trust them the most. I then selected options based on price}', etc. 
In other words, some customers (nearly 1 out of 5) associate trust and reputation with Amazon (the seller) in countries where it sells to customers directly.  
However, nearly 80\% of respondents selected other third-party sellers, showing that they trust other sellers as well.

\noindent
\textbf{Consistency in responses: }To further quantify the relative effect of each individual features on the preference of respondents and see if the respondents' reasons corroborate with their decisions we fit a random forest regression model.\footnote{\url{https://scikit-learn.org/stable/modules/generated/sklearn.ensemble.RandomForestRegressor.html}}
The independent variables are price (normalised by the minimum price within a competition), average rating of the seller, its percentage of positive feedback, its number of ratings and the delivery option (FBA or Not). Given these features, the model is trained to predict the number of times respondents give the first preference votes to the corresponding offers from different sellers. 
The feature importance analysis of the random forest regression model trained on the dataset mostly corroborates with the answers provided by respondents in the survey. 
The relative ranking of the different features as calculated by the random forest regression model is summarised in Table~\ref{Tab: FeatureImportanceRegression}.
For India, \#Ratings is found to be the most influential feature, followed by percentage of positive feedback. For USA, relative price and percentage of positive feedback are found to be the top most influential features in our regression analysis which are among the top features mentioned by the respondents too. 
Much like in USA for France, relative price and percentage of positive feedback are found to be the top most influential features and the feature ranking corroborates with the explanations provided by the respondents too. But in case of Germany, although more people mentioned price in their responses, the random forest analyses shows the relative importance of percentage of positive feedback is higher.
\color{black}

\begin{table}[!t]
	\noindent
        \footnotesize
	\centering
	\begin{tabular}{ |p{1 cm}||p{2.5 cm}|p{2.5 cm}|p{2.5 cm}|p{2.5 cm}|}
		\hline
		{\bf Rank} & \textbf{India}&{\bf USA} & {\bf Germany} & {\bf France}\\
		\hline \hline
		{\bf 1st } & \#ratings & relative price & \% of positive feedback & relative price\\
		\hline
		{\bf 2nd} & \% of positive feedback & \% of positive feedback & relative price & \% of positive feedback\\ 
		\hline
		{\bf 3rd } & average user rating & delivery option & \#ratings & delivery option\\ 
		\hline
		{\bf 4th} & relative price & \#ratings & average user rating & average user rating\\ 
		\hline
		{\bf 5th } & delivery option & average user rating & delivery option & \#ratings\\
		\hline
	\end{tabular}	
	\caption{\textbf{Feature importance ranking in respondents' first preference votes as observed in the survey response based on a random forest regression model. We observe that \% of positive feedback (a seller performance metric) is among the top-2 most influential feature across all the countries. While in India \#ratings is found to be most important, the importance of relative price is higher in the rest of the three countries.}}
	\label{Tab: FeatureImportanceRegression}
	\vspace{-8mm}
\end{table}

\vspace{1mm}
\noindent {\bf Takeaways from this section:}
The primary takeaway from this section are as follows:

\noindent
$\bullet$ Some customers associate trust and / or reputation with Amazon, the seller in countries where Amazon sells on its marketplaces. Yet, in a significant fraction of cases most customers opt for offers from sellers other than Amazon SMs. 

\noindent
$\bullet$ In contrast, Amazon's \buybox{} algorithm pre-selects Amazon SMs in a significantly higher frequency for similar offers as per our dataset.

\noindent
$\bullet$ In India and USA, customers prefer options from Amazon fulfilled sellers (i.e., when the offer is fulfilled by Amazon) more than in European countries like Germany or France.

\noindent
$\bullet$ Lastly, the self reported explanations and a subsequent analyses on random forest regression model suggest that customers give more importance to the reported seller performance metrics -- specifically, (1)~\% of positive feedback, and (2)~number of ratings (primarily in India) -- and the price on the offer listing pages. 

%% file: strike.tex
\section{Striking-through of seller feedback on Amazon}
\label{Sec: StrikeThrough}
Given the importance of seller performance metric e.g., percentage of positive feedback, average user rating etc., in this section we focus on these seller performance metrics and how they are evaluated for each individual seller.

\subsection{Dataset collected for seller feedback}
We collected the seller feedback garnered by 1000 active sellers on {\tt Amazon.in}, {\tt Amazon.com}, {\tt Amazon.de} and {\tt Amazon.fr}. We extracted the set of sellers from a list curated by Marketplace Pulse~\cite{MarketPlacePulsel2022Top}, where the sellers are sorted as per the number of feedback received in last 30 days. 
We collected the seller feedback given to these sellers i.e., the rating (out of 5 stars), the associated review text for each seller feedback (as shown in Figure~\ref{Fig: SellerProfilePage}) and if it was struck through by Amazon or not. The data collection methodology is detailed in Appendix~\ref{Sec: DatacollectionSellerFeedback}. 
Our dataset consists of over 4M reviews received by 4000 Amazon sellers in the last 12 months across the four countries. Our set of 4000 sellers includes 3120 Amazon Fulfilled sellers, 546 Amazon shipped sellers and 334 Merchant Fulfilled sellers. A brief overview of the dataset collected is shown in Table~\ref{Tab: FeedbackDataset}. The numbers in the table should be read as follows: for Amazon fulfilled sellers in the Indian marketplace (first column), the total number of reviews collected was 462,443. Out of these 131,664 were low rated reviews i.e., reviews where the associated rating was 3 stars or below. Note that these are the potential candidates for review striking through since such striking through would ensure that the average user rating and the percentage of positive feedback is unaffected for the seller and is therefore advantageous for the seller. Finally, 97,603 out of the low rated reviews were actually struck through by Amazon.

\begin{table}[!t]
	\noindent
	\scriptsize
	\centering
	\begin{tabular}{ |p{2.0 cm}|c|c|c|c|c|c|c|c|c|c|c|c|}
		\hline
		& \multicolumn{3}{c|}{\textbf{India}} &\multicolumn{3}{c|}{\textbf{USA}} & \multicolumn{3}{c|}{\textbf{Germany}} &\multicolumn{3}{c|}{\textbf{France}}\\
            \hline
           Number of & {\bf AF} & {\bf AS} & {\bf MF} & {\bf AF} & {\bf AS} & {\bf MF}& {\bf AF} & {\bf AS} & {\bf MF}& {\bf AF} & {\bf AS} & {\bf MF}\\
            \hline
           \textbf{sellers }& 693 & 260 & 47 & 900 &33 & 67 & 745 &165 & 90 & 782 &88 & 130\\
            \hline
            \textbf{reviews} & 462,443 & 63,869 & 19,327 & 439,088 & 16,665 & 31,490 & 1,404,880 & 356,565 & 170,795 & 779,471 & 125,931 & 194,700\\
		\hline
            \textbf{Low rated} & 131,664 & 8,909 & 2,978 & 57,567 & 2,038 & 3,262 & 102,774 & 33,908 & 12,944 & 65,284 & 15,762 & 18,452\\
            \hline
            \textbf{struck-through} & 97,603 & 2,001 & 0& 47,287 & 304 & 0 & 57,723 & 1,927 & 0 & 42,010 & 570 & 0\\
            \hline
            \textbf{\% struck-through } & \red{74.13} & 22.46 & 0 & \red{82.14} & 14.92 & 0 & \red{56.16} & 5.68 & 0 & \red{64.35} & 3.62 & 0 \\
            \textbf{(out of low rated) } & &  &  &  &  &  &  &  &  &  &  &  \\
		\hline
	\end{tabular}	
	\caption{\textbf{Statistics of seller feedback data collected for top-1000 active sellers across four of Amazon's marketplaces. We observe that more than half of the low rated feedback that Amazon fulfilled sellers garner are eventually struck through in all of its marketplaces.}}
	\label{Tab: FeedbackDataset}
	\vspace{-7 mm}
\end{table}

\subsection{RQ3 (a): Impact of striking-through negative feedback}\label{Sec: ReEvaluation}
As discussed in Section~\ref{Sec: StrikePolicy}, Amazon has policies in place that allows its \SpSeller{} to apply for striking through some of the low rated seller feedback that they receive. As per Amazon's policies, this striking through also will not reflect on the seller performance metrics.
Table~\ref{Tab: FeedbackDataset} shows that nearly 28\% of the feedback collected for Amazon fulfilled sellers had a low rating associated with it for its Indian marketplace in our dataset. 
Out of these, 74.13\% were struck-through by Amazon (97,603 low-rated reviews were struck through out of 131,664, thus 74.13\%). 
If we consider the percentage of low rated reviews across the different countries, the performance of Amazon fulfilled sellers in general improves in USA, Germany and France (28\% low rated reviews in India as compared to 13\%, 7\% and 8\% low rated reviews respectively for the other countries).
However, out of the low rated reviews 82.14\%, 56.16\%, and 64.35\% reviews were struck through for USA, Germany and France respectively. Sellers who use Amazon's delivery services for shipping their product only (i.e., Amazon shipped (AS) type sellers) has very comparable performance to that of Amazon fulfilled sellers across countries. 
However, the percentage of negative reviews being struck through reduces to 22.46\%, 14.92\%, 5.68\%, 3.62\% (out of all low rated reviews) in the respective countries. For the Merchant-fulfilled sellers, who fulfill their order \textit{without} any assistance of Amazon, no low rated reviews were struck through as this functionality is unavailable for these group of sellers.

\begin{figure}[t]
	\centering
	\begin{subfigure}{0.38\columnwidth}
		\includegraphics[width= \textwidth, height=4cm]{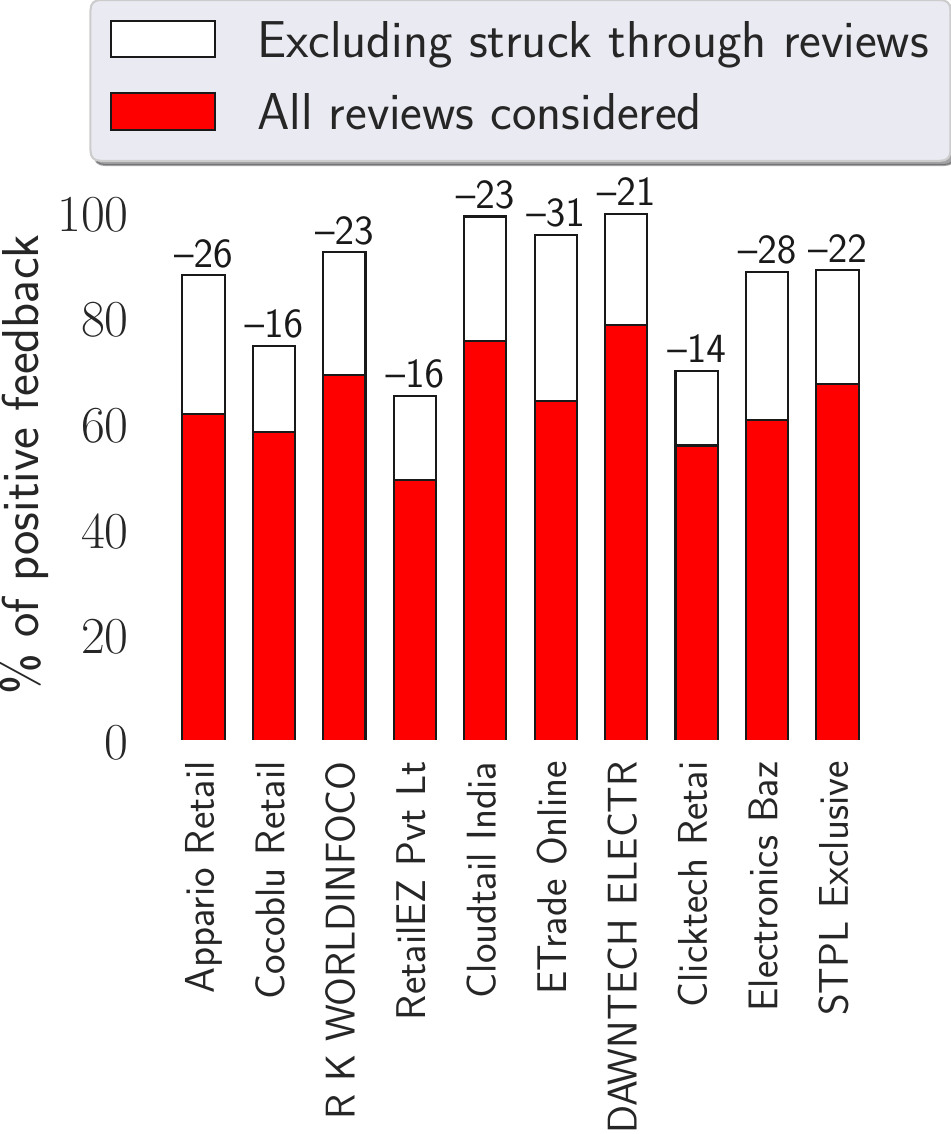}
		\vspace*{-4mm}
		\caption{India}
	\end{subfigure}%
	\begin{subfigure}{0.38\columnwidth}
		\includegraphics[width= \textwidth, height=4cm]{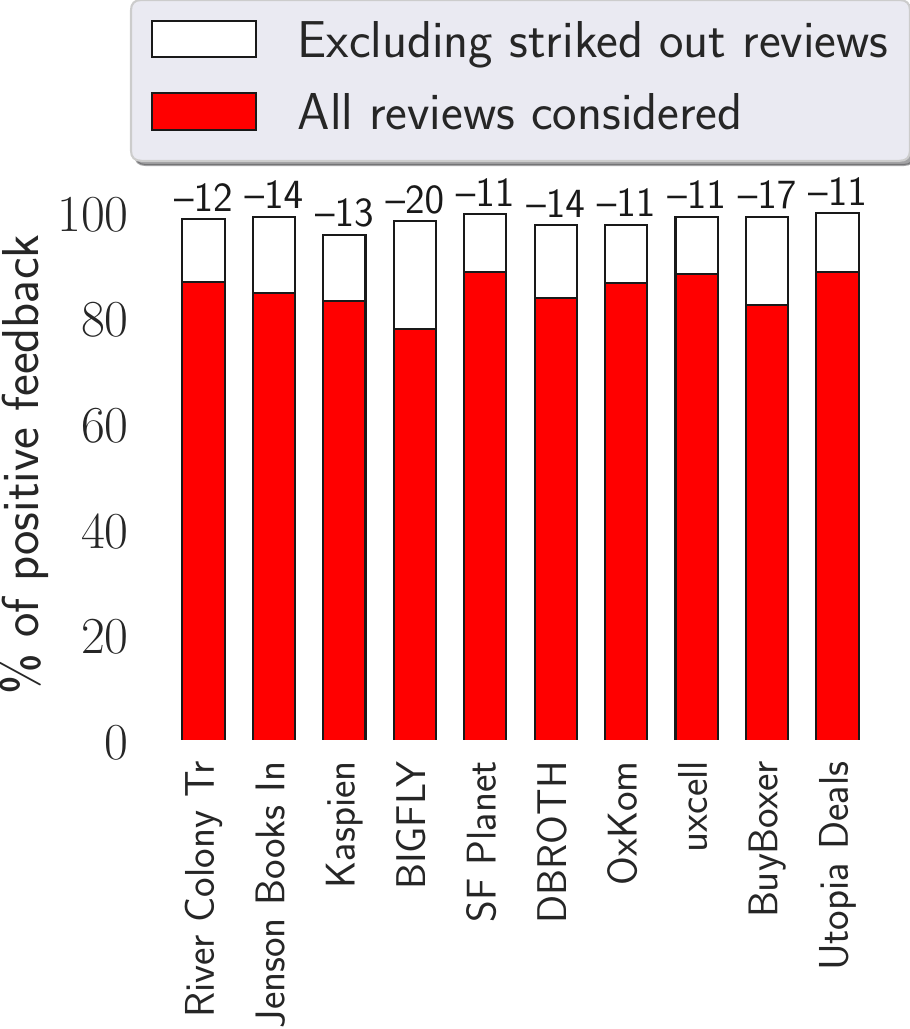}
		\vspace*{-4mm}
		\caption{USA}
	\end{subfigure}

        \begin{subfigure}{0.38\columnwidth}
		\includegraphics[width= \textwidth, height=4cm]{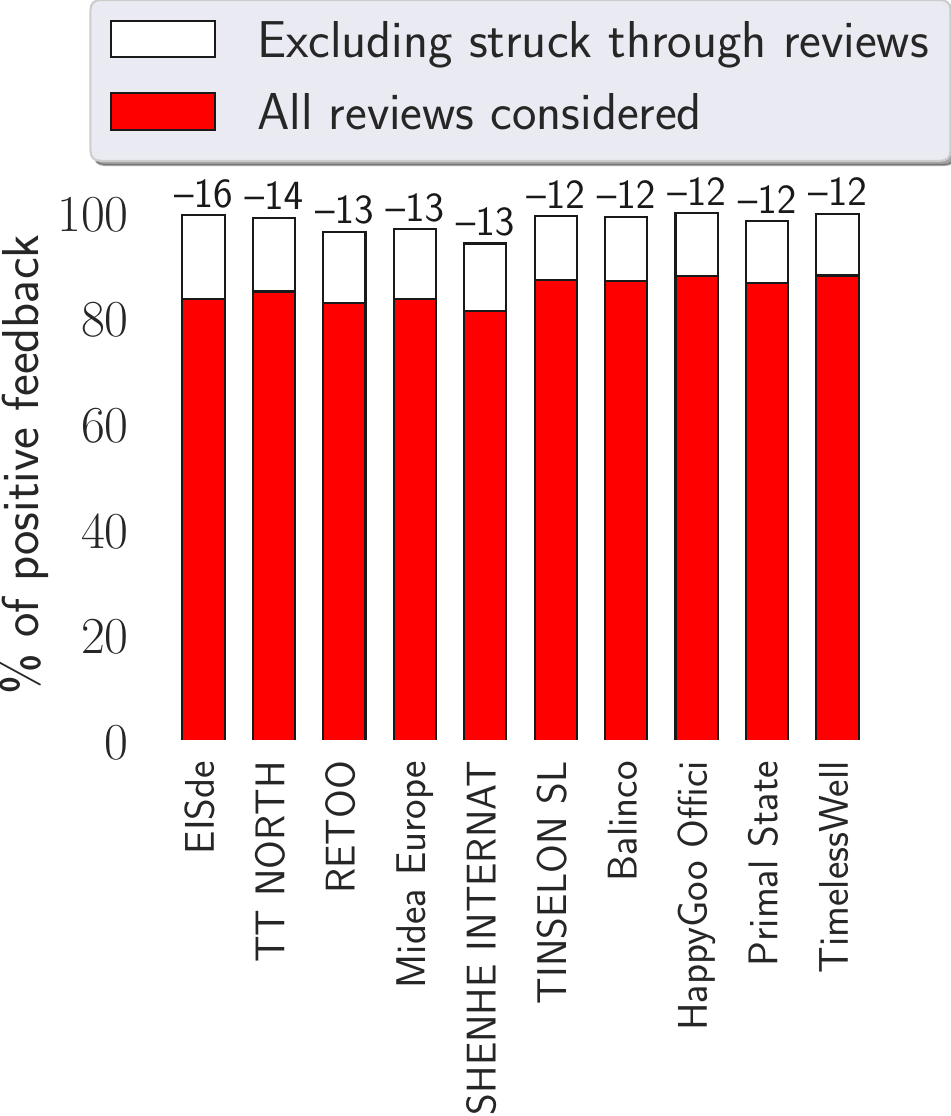}
		\vspace*{-4mm}
		\caption{Germany}
	\end{subfigure}%
	\begin{subfigure}{0.38\columnwidth}
		\includegraphics[width= \textwidth, height=4cm]{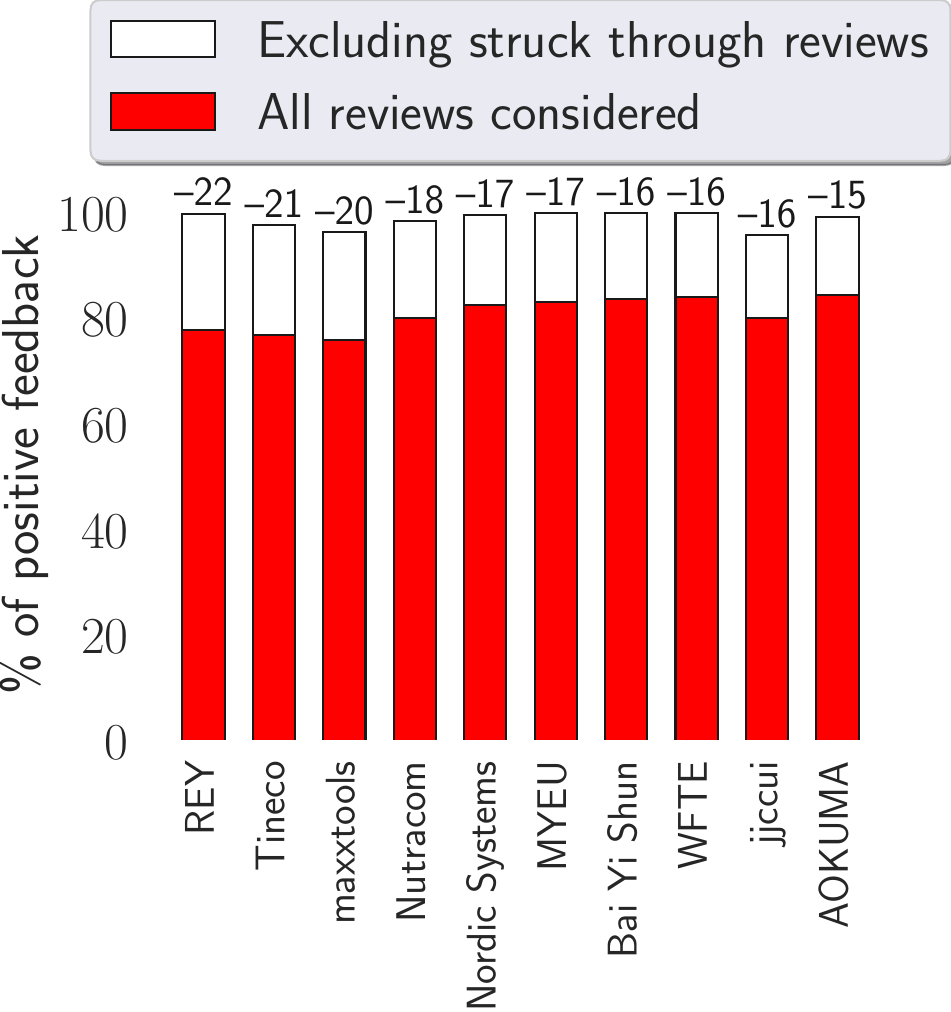}
		\vspace*{-2mm}
		\caption{France}
	\end{subfigure}

	\vspace{-3mm}
	\caption{{\bf Discrepancy in the \% of +ve feedback of some of the top sellers on Amazon's (a)~India, (b)~USA, (c)~Germany, and (d)~France marketplaces. 
    All these sellers have special relationship with Amazon. Amazon significantly over-represents performance metrics for sellers having special relationships.}
	}
	\label{Fig: DiscrepancyPerformance}
	\vspace{-6 mm}
\end{figure}

\vspace{1mm} \noindent
\textbf{Discrepancy in the performance metric of \SpSeller{}:}
Figure~\ref{Fig: DiscrepancyPerformance} shows the percentage of positive feedback calculated with and without the struck-through reviews for some of the top active sellers across the four Amazon marketplaces. The height of the bar for each seller denotes the metric reported by Amazon; while the height of the red portion of each bar is the calculated score by considering all the rating scores (including those which are struck-through). The number on top of each bar denotes the magnitude of drop / increase in the metric. For example, the first seller in Figure~\ref{Fig: DiscrepancyPerformance}(a), Appario Retail (an Amazon SM) reports a \% of positive feedback of 87\% over the past 12 months. However, when we calculate the same including the reviews which have been struck-through, the metric drops to 61\%, i.e., a 26\% drop.
Notice, all of the sellers shown in the figure are Amazon fulfilled and are among the largest active sellers on the respective marketplaces. 

\begin{figure}[t]
	\centering
	\begin{subfigure}{0.24\columnwidth}
		\includegraphics[width=\textwidth, height=3cm]{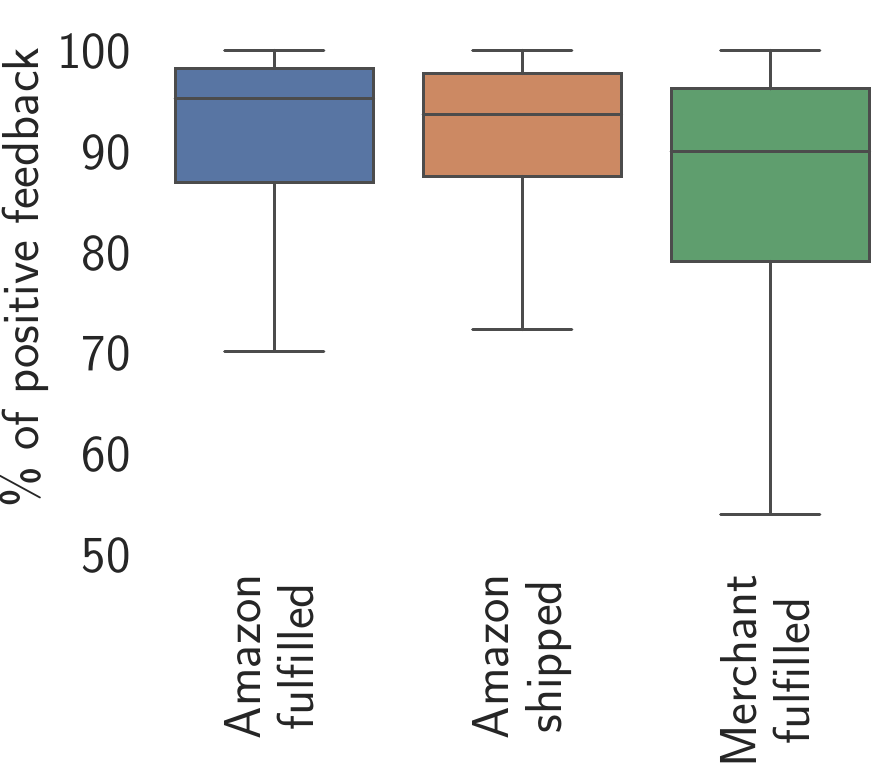}
		\vspace*{-4mm}
		\caption{\footnotesize India: Excluding}
	\end{subfigure}%
	\begin{subfigure}{0.24\columnwidth}
		\includegraphics[width=\textwidth, height=3cm]{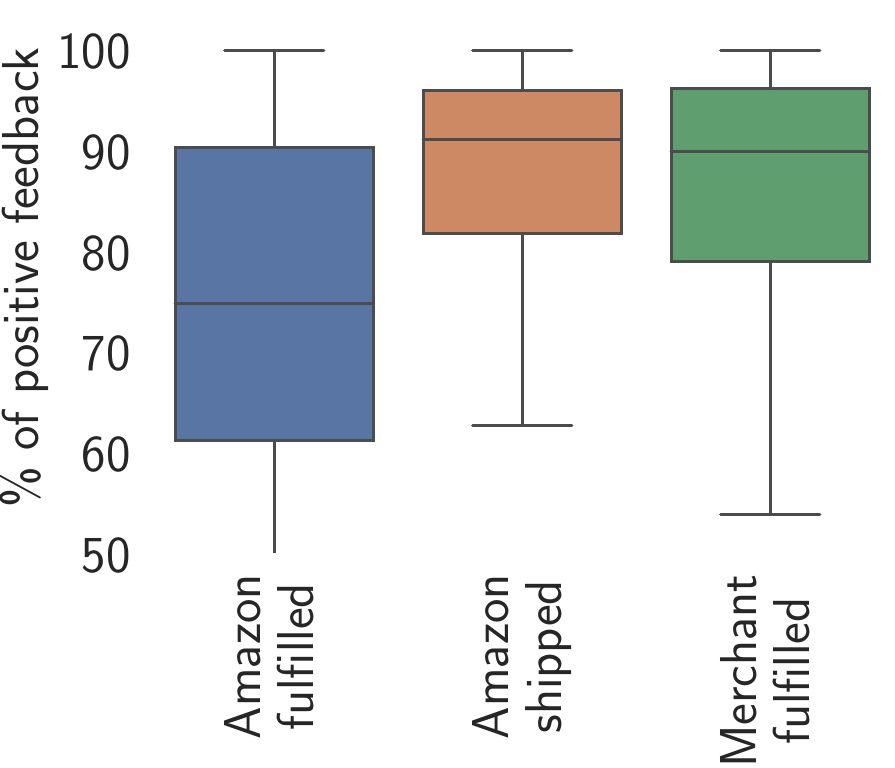}
		\vspace*{-4mm}
		\caption{\footnotesize India: Including}
	\end{subfigure}
        \begin{subfigure}{0.24\columnwidth}
		\includegraphics[width=\textwidth, height=3cm]{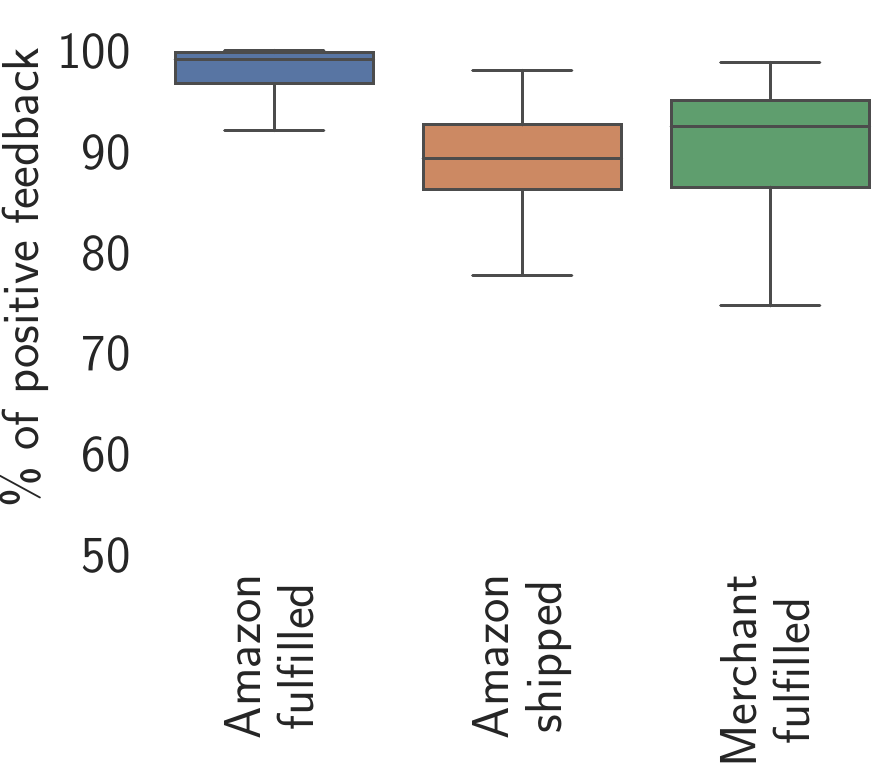}
		\vspace*{-4mm}
		\caption{\footnotesize USA: Excluding}
	\end{subfigure}%
	\begin{subfigure}{0.24\columnwidth}
		\includegraphics[width=\textwidth, height=3cm]{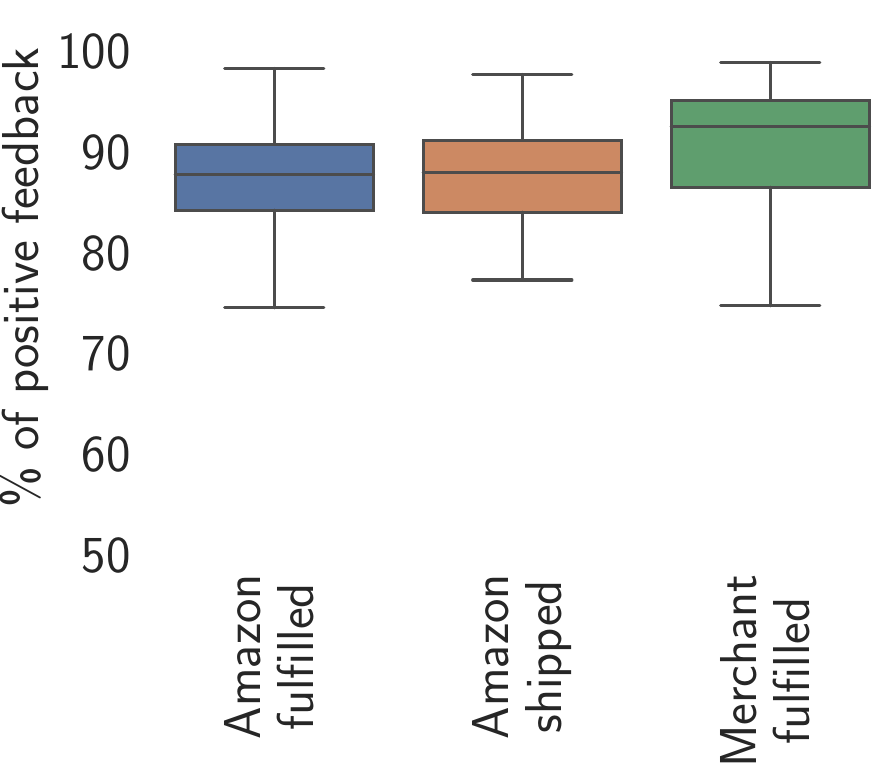}
		\vspace*{-4mm}
		\caption{\footnotesize USA: Including}
	\end{subfigure}

        \begin{subfigure}{0.24\columnwidth}
		\includegraphics[width=\textwidth, height=3cm]{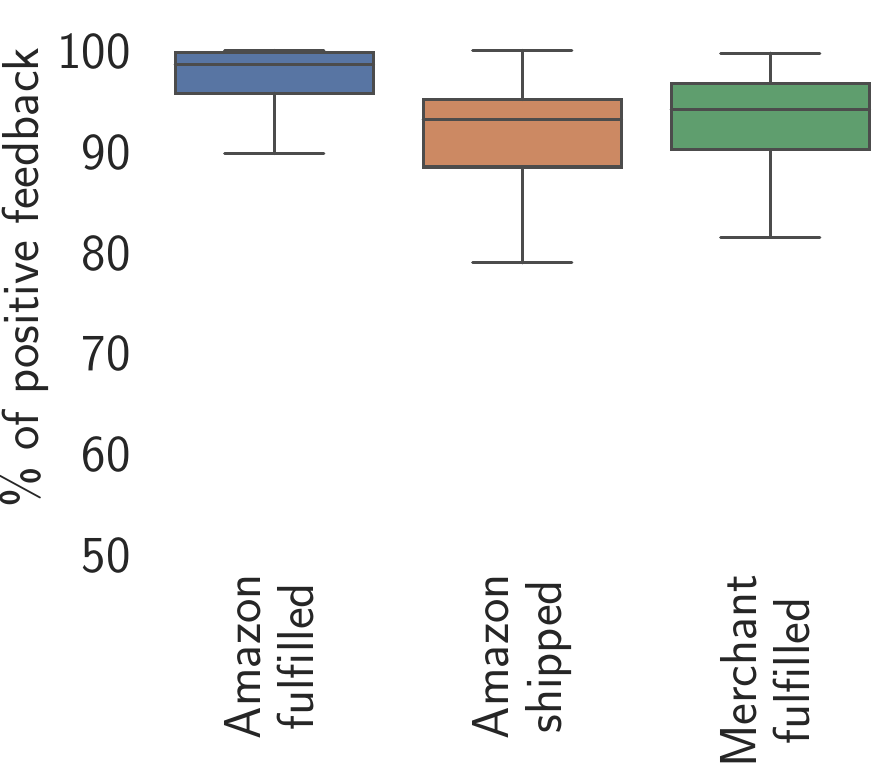}
		\vspace*{-4mm}
		\caption{\footnotesize Germany: Excluding}
	\end{subfigure}%
	\begin{subfigure}{0.24\columnwidth}
		\includegraphics[width=\textwidth, height=3cm]{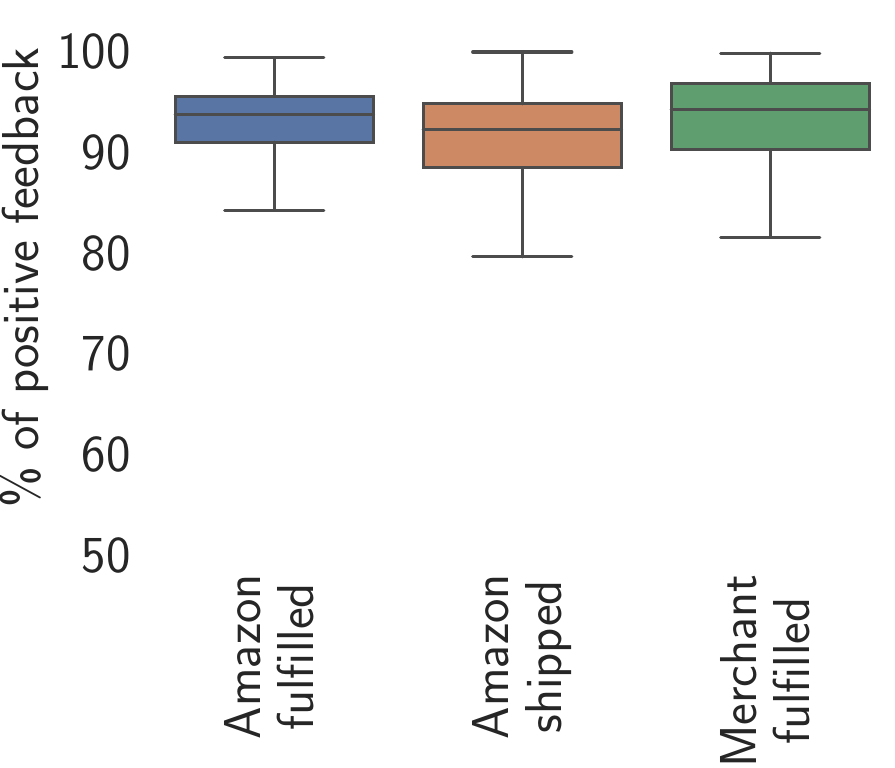}
		\vspace*{-4mm}
		\caption{\footnotesize Germany: Including}
	\end{subfigure}
        \begin{subfigure}{0.24\columnwidth}
		\includegraphics[width=\textwidth, height=3cm]{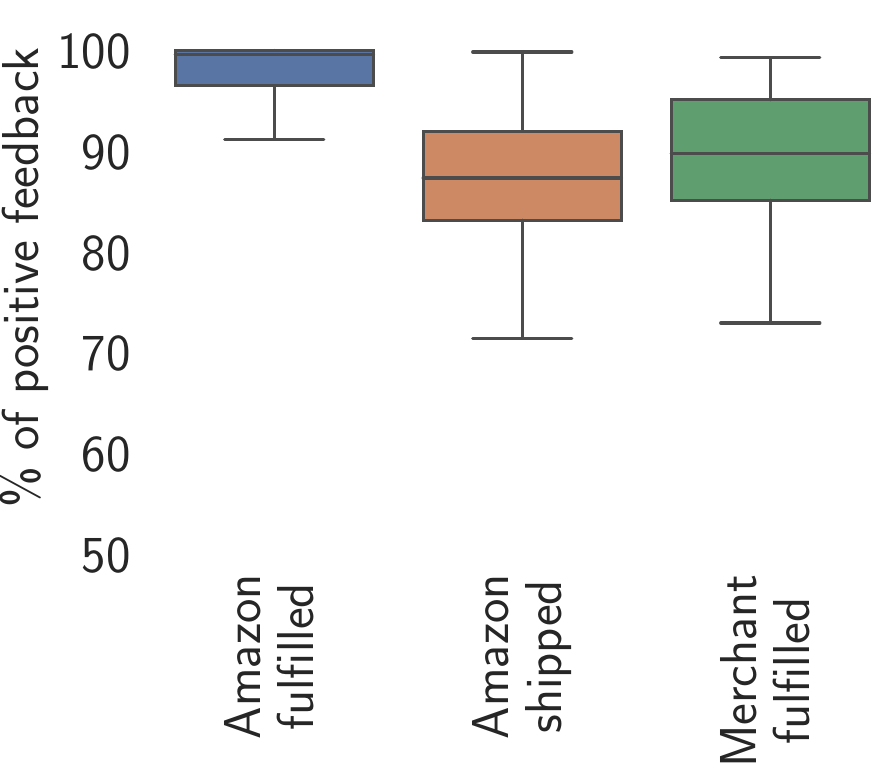}
		\vspace*{-4mm}
		\caption{\footnotesize France: Excluding}
	\end{subfigure}%
	\begin{subfigure}{0.24\columnwidth}
		\includegraphics[width=\textwidth, height=3cm]{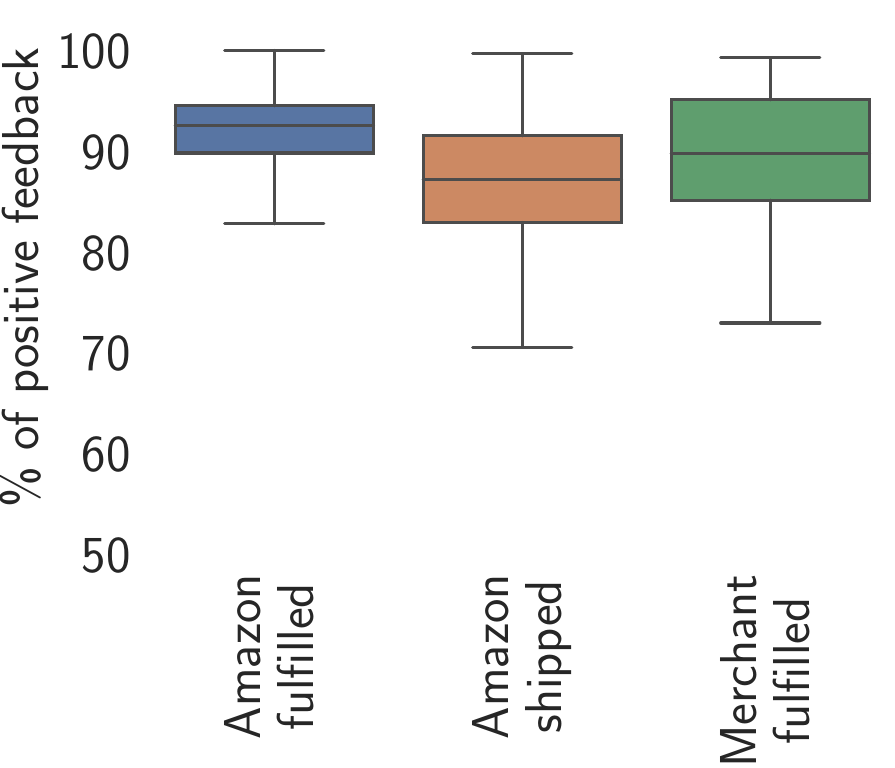}
		\vspace*{-4mm}
		\caption{\footnotesize France: Including}
	\end{subfigure}
	\vspace{-3mm}
	\caption{{\bf \% of positive feedback of AF, AS and MF sellers calculated excluding and including struck-through reviews for the different Amazon marketplaces. The distribution of AF sellers are very different (as shown in the first box plots in the consecutive figures) highlighting the effect of review striking through on seller performance metrics of \SpSeller{}.}
	}
	\label{Fig: ImpactofStriking}
	\vspace{-3mm}
\end{figure}

Next, we perform the analyses with sellers grouped together in categories based on how the handle their delivery logistics. 
Figure~\ref{Fig: ImpactofStriking} shows the box plot of \% of positive feedback of all the sellers in the different groups. In Figure~\ref{Fig: ImpactofStriking}(a, c, e and g), we show the distribution of the \% of positive feedback of the sellers as reported by Amazon (i.e., evaluated excluding the struck-through reviews). In Figure~\ref{Fig: ImpactofStriking}(b, d, f and h), we show the distribution of the \% of positive feedback of the sellers evaluated including the struck-through reviews. 
Notice, the distributions where we exclude the struck through reviews suggest the \% of positive feedback and hence the quality of service of Amazon fulfilled sellers are the best among the three types. 
However, as soon as we include the struck through reviews in the evaluation, the median percentage of positive feedback for Amazon fulfilled sellers drops by 20\%, 11\%, 5\%, and 7\% for India, USA, Germany and France respectively. In fact, for India, USA and Germany, the median value for the Amazon fulfilled sellers actually drops below that of Merchant fulfilled sellers. 
In addition, a students' $t$-test confirmed that all these drops were statistically significant.
In other words, if not for the review strike-through policy of Amazon, the percentage of positive feedback of merchant fulfilled sellers are in fact better than that of Amazon fulfilled sellers, across several countries.

Note that adaptation of such differential policies based on special relationship with the marketplace is in direct violation of the prescription of the Digital Markets Act (see Section~\ref{Sec: StrikePolicy} for more details). At the same time, our experiment shows that the review strike through policy improves the seller performance metric of the \SpSeller{} (esp. Amazon fulfilled sellers) significantly. 
Given the importance that seller performance metrics have in the decision making process of both customers (as discussed in Section~\ref{Sec: SurveyBuyBox}) and in the \buybox{} algorithm~\cite{Amazon2021Becoming}-- such discrepancy and therefore such policies can potentially result in nudging customers more toward \SpSeller{} as compared to other potentially more deserving third party sellers.

%% file: interpret-Policy-New.tex
\subsection{RQ3 (b): Transparency, interpretability and fairness of strike-through policy}
To understand transparency, customers' interpretability and perception of Amazon's feedback strike-through, we conducted another survey on Prolific. 
\new{We recruited 100 participants on Prolific who are fluent in English and have a very high approval rate ($\ge98\%$) in prior participated surveys. Each respondent was remunerated at a rate of 9 GBP per hour which is recommended by Prolific to be a good and ethical rate~\cite{Prolific2023Payment}.}
The gender distribution was balanced at 50-50 for this survey. Most of the respondents are familiar with Amazon e-commerce platform and regularly shop on Amazon too. Next, we discuss the insights from the survey.

\noindent
\textbf{Transparency of Amazon's strike-through policy:} We ask the respondents to put on the hats of customers of Amazon and answer if they were aware of feedback striking-through being done on Amazon before taking our survey. \textit{78 out of the 100 participants stated that they were not aware of such striking-through on Amazon before taking our survey} (even though they self-report to be regular shoppers on Amazon). This lack of awareness of respondents about this important policy hints at lack of transparency on the part of Amazon marketplace for the strike-through policy they adopt for evaluating the seller performance metrics. 
In fact, one would not be aware of this policy unless he/she goes through the seller profile page of some of the sellers where they will be able to see seller feedback as shown in Figure~\ref{Fig: SellerProfilePage}.

\noindent
\textbf{Interpretation of Amazon's strike-through policy: }
In the second part of the survey we ask our respondents to evaluate if the policy is being implemented properly. Notice Amazon mentions if the entire comment relates to \textbf{\textit{delivery experience}} then Amazon may strike-through such reviews for Amazon fulfilled sellers. 
`Delivery experience'-- is an umbrella term which may include an entire array of issues. 
We came up with eight distinct types of issues, from a manual inspection of a random sample of struck-through reviews. 
These issues are -- 
(i)~order not delivered, (ii)~other delivery issues, (iii)~wrong product delivered, (iv)~damaged product delivered, (v)~return related issues, (vi)~refund related issues, (vii)~product quality related issues, (viii)~canceled order.
We wanted to understand which issue (out of these eight) respondents deem to be within the scope of `Delivery experience' and to what degree. 
To this end, we randomly selected 30  struck-through reviews written in English received by Amazon fulfilled sellers. These reviews correspond to different issues as mentioned above. 

\new{During the survey, we show Amazon's strike-through policy along with a brief description about the ``Fulfilled by Amazon'' concept to the survey participant. Then we ask the following question:\\
\textit{Review: ``$<$Text of the struck-through review$>$''}\\
\textit{Amazon has struck-through the above review. Do you think it should have been struck-through as per Amazon's review strike-through policy?}\\
Further, a follow-up question was also asked as to \textit{why} they think a particular review should / should not be struck-through.}

\begin{figure}[t]
	\centering
	\begin{subfigure}{0.45\columnwidth}
		\includegraphics[width=0.8\textwidth, height=4cm]{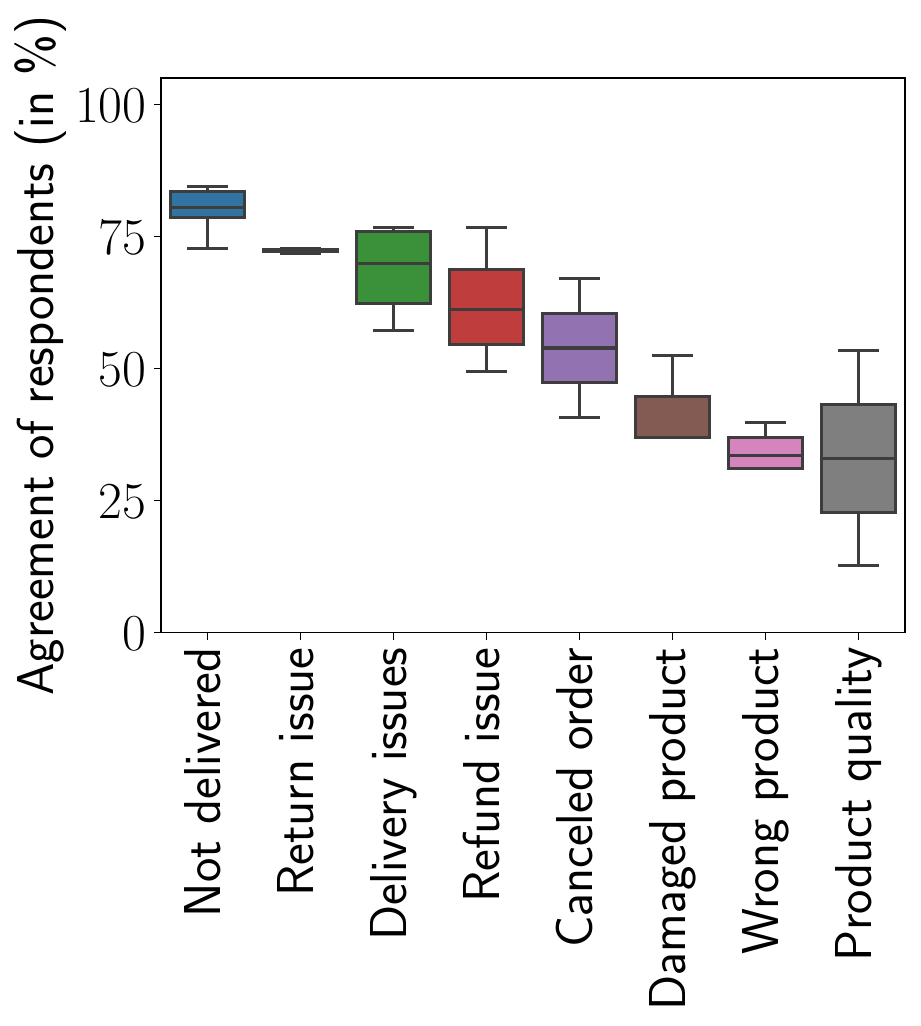}
	\end{subfigure}%
	\vspace{-3mm}
	\caption{{\bf \final{Percentage of respondents who agreed that reviews mentioning the corresponding issue should be struck-through as per Amazon's policy. The agreement is high for delivery-related issues, but very low for issues mentioning wrong products delivered or product quality.}}
	}
	\label{Fig: humanCodes}
	\vspace{-5mm}
\end{figure}

\noindent
\textbf{Observation:} We received a total of 3000 evaluations for the 30 struck-through reviews. 
In 1826 cases respondents agreed that the corresponding reviews should have been struck-through as per Amazon's policy (as it is worded currently). 
\textit{In the remaining 39\% of cases respondents did not agree with Amazon striking-through a particular review.}
Table~\ref{Tab: ReviewResponse} shows examples of reviews that have been struck-through by Amazon, but majority of our respondents felt that these reviews should \textit{not} have been struck-through under the current Amazon policy.

\begin{table}[!t]
	\noindent
	\scriptsize
	\centering
	\begin{tabular}{|p{6 cm}||p{7.5 cm}|}
		\hline 
		\textbf{struck-through reviews for AF sellers} & \textbf{Reasons why they should not be struck-through (as per respondents)} \\
		\hline
		We returned this item and never received any refund. The refund has been pending for over a month now. Very bad service. & Others should be aware of such potential risks. \\
		\hline
		Even the replacement I received is wrong, you have mentioned c-type cable but sending it an a-type cable. & It was sent a total different product \\
		\hline
		Product was so low quality. & The seller is accountable for the quality of products\\
		\hline
		Why are you showing here different items and sending some different cheats to customers? & the seller is responsible for advertisement of products in stock or being sold\\
		\hline
		Ordered 4 pcs of Milton 1800ml glass jars. Received one damaged 1000ml. Absolute cheating by the seller. Requested for replacement with the ordered item. Still I am waiting for it. & This is an honest review about the seller that could help other clients so it shouldn't have been taken down \\
		\hline
	\end{tabular}	
	\caption{ \textbf{Reviews received by Amazon fulfilled sellers which were struck-through by Amazon. However, majority of the respondents in our survey thought these reviews should not be struck-through due to the mentioned reasons.}}
	\label{Tab: ReviewResponse}
	\vspace{-6 mm}
\end{table}

The agreement of annotators with the Amazon's strike-through policy varied across the different issues that the reviews mention about. 
Figure~\ref{Fig: humanCodes} shows the distribution of percentage of agreement (that a review should have been struck-through) of annotators for different issues that we observed in the random sample. 
The average agreement rate, i.e., \% of respondents who agreed with the review being struck-through, was higher for `not delivered' (80\%), `return issue' (72\%), and `delivery issues' (68\%). For some of the other issues such as `product quality'  (33\%), `wrong product delivered' (34\%), and `damaged product delivered' (42\%) the agreement was lower. 
We include some of the reasons provided by respondents in Table~\ref{Tab: ReviewResponse} too.
These observations highlight the gap between the way the policy is being implemented by Amazon today, and its interpretation by the respondents.

\noindent
\textbf{Fairness of Amazon's strike-through policy: }
Our next set of questions to the respondents was to think normatively and answer whether they agree with the strike-through policy in place by Amazon. While 45 respondents disagreed with Amazon's strike-through policy, 47 respondents agreed to the policy (remaining participants being neutral). Thus, respondents are in a split about whether the policy in place is agreeable or not. 
We further asked them whether such strike-through policies should be uniform across all sellers irrespective of their fulfillment service. \textit{69 respondents answered in the affirmative}.
Thus, majority of the respondents think that if such policies are to be in place, they should be uniformly applicable to all sellers irrespective of their special relationships.

\vspace{1mm}
\noindent {\bf Takeaways from this section:}
To summarize the major takeaways from this section,

\noindent
$\bullet$ Amazon's review strike-through policy has significant impact on the evaluation of performance metrics of \SpSeller{}, 
resulting in high discrepancy between the shown and the actual metrics.

\noindent
$\bullet$ 78\% of the respondents are not aware of striking-through of negative reviews on Amazon, highlighting the lack of transparency on this matter. 

\noindent
$\bullet$ Given Amazon's policy, respondents do not agree with reviews discussing certain issues should be struck-through. These observations indicate that Amazon may be trying to favour its \SpSeller{} by brushing many factors under the `carpet' of feedback strike-through.

\noindent
$\bullet$ 69\% respondents think that if at all such policies are in place, they should be in place uniformly for all sellers irrespective of the fulfillment services they opt for. 

%% file: sellerMetric.tex
\section{RQ3 (c): Impact of performance metrics on customers' decisions}~\label{Sec: SurveyBuyBoxWithoutRateNum}
In the previous sections, we showed how different seller performance metrics influence the customers' decision-making while choosing among different offers. 
Sometimes they prioritise features like \#Ratings, which is an indicator of the scale at which a seller operates, or features like percentage of positive feedback, whose accuracy is questionable given Amazon's review strike-through policy. 
To understand and quantify the extent of the influence of these metrics on the customers, we conducted some more surveys similar to the survey presented in Section~\ref{Sec: SurveyBuyBox}, with several changes in the settings (which we shall describe next). 
Note that although we have discussed about what features influence customers' decision making according to their self-reported explanations and a regression study, it is important to quantify the exact effects by putting customers in a counterfactual setup e.g., where \#Ratings is not shown or where rectified seller performance metrics (including the strike through reviews) are shown etc. Such a setting may bring out the precise quantifiable effect of the changes in the customers' decision-making.
\new{We chose to conduct these additional surveys on the Indian marketplace, mainly because we have access to the performance metrics of all the sellers only for the Indian marketplace. Recall that, in the other countries, Amazon does not show any performance metric for itself, nor does it have any seller profile page, which essentially hinders  conducting this study in the other marketplaces.} 

\new{We now describe the additional surveys that we conducted. Each setting, we discuss below, is shown to 50 Indian nationals with approval rate greater than 98\% on Prolific. Again, all the participants were remunerated at a rate of 9 GBP per hour.
57\% participants in the survey reported their gender to be male while the remaining 43\% reported to be females. 72.5\% of the participants self reported to be frequent shoppers on Amazon while 19\% reported they moderately shop on Amazon too. The four groups of 50 participants are mutually exclusive to one another.}

\vspace{1 mm}
\noindent
\textbf{Control setting (Amazon metrics w \#Ratings): }This is the exact survey conducted in Section~\ref{Sec: SurveyBuyBox} for the Indian marketplace where we had shown 15 products and the top-4 offers to 50 respondents from India. To remind the readers about our questions, we asked the respondents ``Suppose you are willing to buy a
`<product name>' on Amazon and these are the offers from different sellers for the same product. Which one would you prefer to buy?''. Further to understand about what feature is important for the customers to make their decision, we also asked them the following: ``Briefly reason about the order of your preferences.''

\vspace{1 mm}
\noindent
\textbf{Treatment 1 (Amazon metrics w/o \#Ratings): }We pose the exact same questionnaire for the same 15 products, only omitting the \#Ratings feature for all offers. Notice the other performance metrics such as average user ratings, percentage of positive feedback were visible to the respondents along with the other features as mentioned in Section~\ref{Sec: SurveyBuyBox}. Note that this survey will indicate the effectiveness of \#Ratings feature on the decision making of customers.

\vspace{1 mm}
\noindent
\textbf{Treatment 2 (Rectified metrics w/o \#Ratings): }We repeat the survey for the same 15 products, this time with the rectified performance metrics. To this end, we collect data from the seller profile page of the sellers involved in the surveyed competitions and repeat the process mentioned in Section~\ref{Sec: ReEvaluation} to evaluate their performance metrics including all the struck through reviews. Now the respondents will see the newly evaluated performance metrics (which include the struck through reviews). 
Notice, this setting will help in quantifying the effects of review strike-through policies of Amazon on the decision making of customers.

\vspace{1 mm}
\noindent
\textbf{Treatment 3 (Rectified metrics with \#Ratings): }Finally, we repeat the survey for the same 15 products, this time with the rectified performance metrics along with the \#Ratings shown. During the survey in Section~\ref{Sec: SurveyBuyBox}, we observed that \#Ratings was mentioned to be highly influential in the Amazon India setting. Thus, to conclude the loop in this setting we show the respondents the rectified metrics as evaluated for Treatment 2, but now with an additional \#Ratings feature.

\begin{figure}[t]
	\centering
	\begin{subfigure}{0.7\columnwidth}
		\centering
		\includegraphics[width= 0.7\textwidth, height=4.5cm]{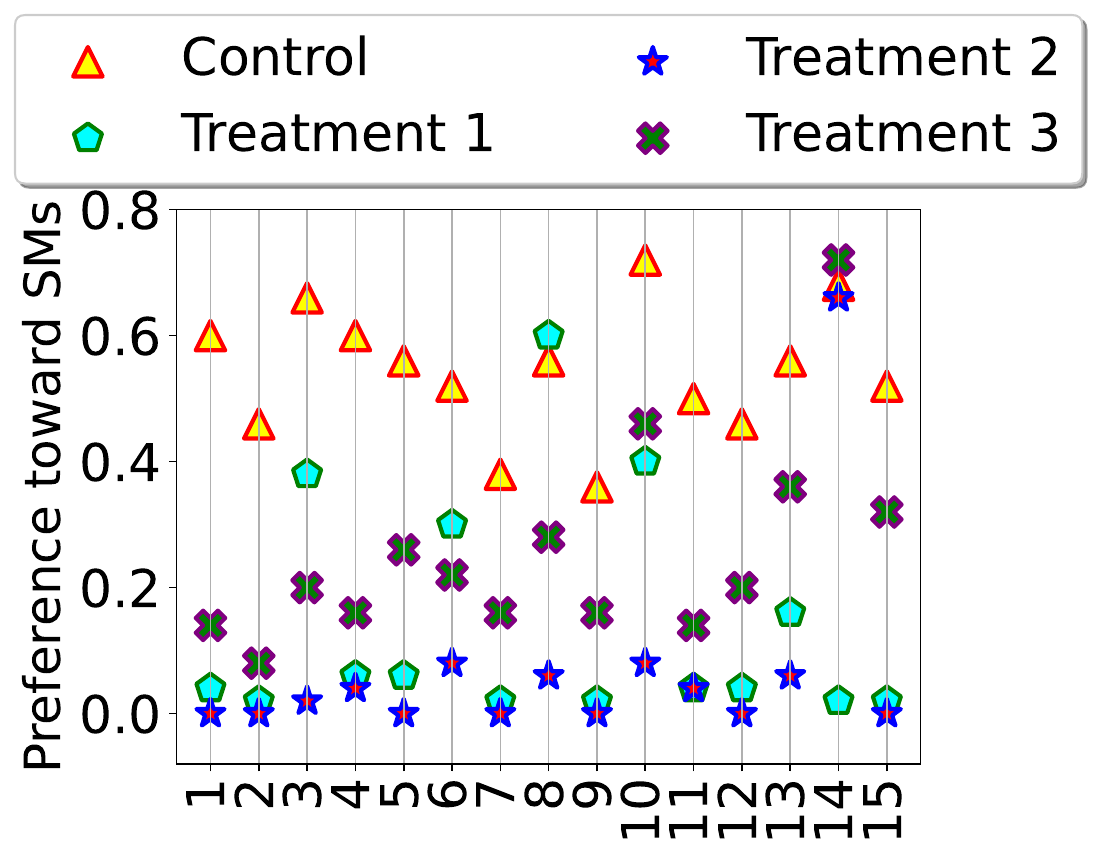}
		\vspace*{-2mm}
	\end{subfigure}%
	
	\vspace{-3mm}
	\caption{{\bf \final{Fraction of first preference votes to Amazon SMs in control and all the treatment settings as per the different products. Showing the rectified metrics and omitting the \#Ratings feature has significant effect on respondents' first preference votes.}}
	}
	\label{Fig: ImpactofFeatures}
	\vspace{-4 mm}
\end{figure}

\begin{table}[!t]
	\noindent
    \footnotesize
	\centering
	\begin{tabular}{ |p{3.5 cm}|p{2.5 cm}|p{2.5 cm}|}
		\hline
		{\bf \% of 1st Preference Votes} & {\bf Amazon Metrics} & {\bf Rectified Metrics}  \\
		\hline \hline
        \multicolumn{3}{|c|}{\textbf{Special Merchants}}\\
        \hline
		  With \#Ratings & 54.26\%  & 28.53\% \\ 
		\hline
		Without \#Ratings & 14.53\% & 6.93\% \\ 
		\hline \hline
        \multicolumn{3}{|c|}{\textbf{Amazon Fulfilled}}\\
        \hline
		With \#Ratings & 81.20\%  & 64.53\% \\ 
		\hline
		Without \#Ratings & 72.13\% & 54.6\% \\ 
		\hline
	\end{tabular}	
	\caption{{\bf Percentage of first preference votes aggregated by Amazon Special Merchants and Amazon fulfilled sellers in the different survey settings. We observe that the percentages drop when we show rectified metrics as opposed to metrics shown by Amazon or when we omit the \#Ratings feature. }}
	\label{Tab: ImpactofFeatures}
	\vspace{-8mm}
\end{table}

\vspace{1 mm}
\noindent
\textbf{Observation: }
We discuss the effect of the different treatment setting first on the Amazon Special Merchants (Cloudtail and Appario for Amazon.in) and then on the Amazon fulfilled sellers (sellers with Amazon FBA option).
In Figure~\ref{Fig: ImpactofFeatures} we show the fraction of first preference votes given by respondents to the Amazon Special Merchants for the different settings of the survey explained above. 
We observe that, except product number 8, in all the other 14 cases, the preference toward SMs have drastically reduced upon omitting the \textbf{\#Rating} feature (pentagon markers for Treatment~1 are much below the triangle markers for the Control). 
When we show rectified metrics without \#Ratings (i.e., Treatment~2), the percentage of first preference votes reduces further significantly for Amazon SMs in all cases (the star markers for Treatment~2 are lower than triangles and pentagons) except product 14\footnote{Upon closer inspection, we found that the performance metrics of the other Amazon fulfilled sellers for this product also reduced significantly, thereby, keeping the Amazon SM as the seller with best performance metrics.}.
Similarly, across all but one case (in product number 14), the fraction of first preference vote reduces when we show the rectified metrics as opposed to the metrics shown by Amazon excluding the struck-through reviews (the cross markers for Treatment~3 are much below the triangle markers for the Control).

In Table~\ref{Tab: ImpactofFeatures} (upper panel), we show the percentage of first preference votes aggregated by Amazon SMs in the different settings of our surveys. In the control setting, (as reported earlier) 54\% of the first preference votes are given to Amazon SMs. However, in the Treatment 1 setting, we observe that the percentage of first preference votes to Amazon SMs reduces to merely 14.5\% (i.e., 109 first preference votes out of 750). In other words, the removal of \#Ratings feature massively reduced the number of first preference votes going to Amazon SMs (by 39.5\%). In Treatment 2, this percentage further reduces to 6.93\% (i.e., 52 first preference votes out of 750). So, with the rectified metrics the percentage of first preference votes got halved as compared to the numbers in Treatment 1. Finally, when we show the respondents the rectified metrics with the \#Ratings, the percentage of first preference votes increased to 28.53\%, thus again highlighting the effect of \#Ratings. However, if we compare the control setting and the Treatment 3 the drop off is still almost nearly 25.47\%. Notice the only difference between the two settings is that, in the control setting, we show seller performance metrics excluding the struck-through reviews (as done by Amazon) whereas in Treatment 3, we show the rectified performance metrics.

Now, let us consider the differences in the first preference votes toward Amazon fulfilled sellers (i.e., sellers with FBA options including the two Amazon SMs) as shown in Table~\ref{Tab: ImpactofFeatures} (lower panel). In the control setting, 81.2\% of the first preference votes went to AF sellers. This number reduces to 72.13\% when we simply remove the \#Ratings feature. However, when we show the rectified metrics to respondents in Treatment 2 this number reduces to 54.6\%. Upon inclusion of \#Ratings it further increases to 64.53\% but never gets to where it was in the control setting. This may be an artifact of the drop in the Amazon fulfilled sellers' performance metrics (as corroborated in Figure~\ref{Fig: ImpactofStriking}).

Moreover, in contrast to the earlier survey, now \% of positive feedback, average user ratings of sellers were mentioned by respondents when asked about features that influence their decisions (especially in Treatments~2 and 3-- where rectified seller performance metrics were shown). The same is also corroborated by the random forest regression model fit on the new set of first preference votes, wherein the relative importance of \% of positive feedback and relative price has increased significantly as indicated in the last column of Table~\ref{Tab: FeatureImportanceRegression2}.

\begin{table}[!t]
	\noindent
        \footnotesize
	\centering
	\begin{tabular}{ |p{1 cm}||p{2.5 cm}|p{2.5 cm}|p{2.5 cm}|p{2.5 cm}|}
		\hline
		{\bf Rank} & \textbf{Control}&{\bf Treatment 1} & {\bf Treatment 2} & {\bf Treatment 3}\\
		\hline \hline
		{\bf 1st } & \#ratings &  \% of positive feedback & average user rating & \% of positive feedback\\
		\hline
		{\bf 2nd} & \% of positive feedback & relative price & \% of positive feedback & average user rating\\ 
		\hline
		{\bf 3rd } & average user rating & average user rating & relative price & \#ratings\\ 
		\hline
		{\bf 4th} & relative price & delivery option & delivery option & relative price\\ 
		\hline
		{\bf 5th } & delivery option & -- & -- & delivery option\\
		\hline
	\end{tabular}	
	\caption{\textbf{Feature importance ranking in respondents' first preference votes as observed in the survey response based on a random forest regression model. Percentage of positive feedback and average user rating's importance have relatively increased in different treatment settings.}}
	\label{Tab: FeatureImportanceRegression2}
	\vspace{-8mm}
\end{table}

\vspace{1 mm}
\noindent
\textbf{Takeaways from this section:} To summarize the major takeaways from this section,

\noindent
$\bullet$ Although it is \textit{not} a genuine indication of quality of service (rather indicates the scale of operation of sellers), \#Ratings is the feature that largely influences the respondents decision-making. As has been argued earlier, this may nudge customers more toward larger sellers (many of whom are \SpSeller{}), to the disadvantage of smaller third-party sellers.

\noindent
$\bullet$ Showing the rectified performance metrics of the Amazon  \SpSeller{} reduces the percentage of first preference votes that they receive during the survey considerably. This suggests that Amazon's review strike-through policy may potentially have considerable effects on the decision making of customers by altering the seller performance metrics on the offer listing pages.

%% file: discussion.tex
\section{Concluding discussion}

\noindent
\new{\textbf{Summary of the work:}
To our knowledge, ours is the first systematic attempt of an end-to-end investigation on Amazon e-commerce marketplace, for potential preferential treatment toward its \SpSeller{} due to different design choices and policies in place by Amazon.
We studied Amazon's choice architectures deployed to assist customers in cases where multiple sellers compete for selling the same product (over Amazon marketplaces in four different countries). 
The study was conducted on Amazon's \buybox{} algorithm, offer listing page and evaluation of seller performance metrics in the context of its review strike-through policies.
Our empirical analyses on 75K+ \buybox{} competitions showed that Amazon won the \buybox{}es in more than 80\% cases when it competed in our dataset. 
However, upon surveying competitions of 60 products among 200 participants, we observed significant gap between the algorithmic choices and customer choices. During the survey, customers also mentioned that seller performance metrics influence their decision the most. This led us to understand how these seller performance metrics are being evaluated. 
To this end, our empirical analyses over 4M seller feedback reviews collected for 4000 top sellers (across the four Amazon marketplaces) suggests that Amazon adopted different policies for evaluating seller performance metrics for sellers who use Amazon's subsidiary services (e.g., FBA).
Finally, we also conducted surveys where we show participants the rectified metrics and observed significant differences in preference toward Amazon and its \SpSeller{} among customers.}

\subsection{Limitations of the work}

\textbf{Limitations of the methodologies adopted}: \new{Accessibility to the deployed algorithms and policies would have been the best possible way to approach the investigation at hand. However, a primary limitation of the work is our partial knowledge of the features that are actually used in the algorithms deployed by Amazon. It can be noted that this limitation is present in any \textit{third-party} audit of such deployed proprietary systems. The proprietary nature of the algorithms, in fact, led us to the research pipeline that has been adopted in this work. Nevertheless, we have attempted our best to make the analysis as complete as possible using the publicly known features. Furthermore, we have analysed the \buybox{} competitions for a limited set of products for some of the most searched queries which may bring in problems pertaining to selection bias. However, we want to point out that these are products appearing on the first search engine result pages of Amazon's four emerging marketplaces for most popular keywords. Thus, we believe, this strategy enabled us to check for the competition of some of the most sought after and popular product pages on Amazon platform across the different marketplaces.} 

\noindent 
\textbf{Limitations in the survey setting:}
\new{Another potential limitation pertains to our conducted surveys where customers were asked to order their preferences. We understand that providing preference votes and doing an actual online purchase are two very different settings. Due to lack of other more suitable options, we chose to proceed with the design choices as discussed in the paper. Similarly, with regards to `delivery time' feature, we did not include the exact \textit{delivery duration} during our surveys or analyses. Notice, the delivery duration from the same seller and for the same product can also differ at different timestamps while delivery option (FBA or not) is a more stable indicator. Furthermore, most of our respondents were frequent users of Amazon and are aware of faster shipping being a well-known advantage of Amazon FBA. Their delivery option choices reflected their preference toward faster delivery in general too (see Table~\ref{Tab: ImpactofFeatures}). 
Thus, although exact delivery dates were not included in the survey, the usage of the proxy (delivery option) conveys the faster delivery promises. And preference toward the same was also observed during the survey as discussed in the paper.}

\new{While the above limitations may affect the findings quantitatively, we believe the qualitative findings will not be significantly different. 
Moreover, the findings of potential preferential treatment corroborates with similar practices on Amazon ecosystem in other prior studies as well~\cite{Yin2021Search, dash2021when}.}

\subsection{Recommendations and takeaways for important stakeholders}
Since such choice architectures have such far-reaching effects on multiple stakeholders of the marketplace, next we briefly discuss potential recommendations and takeaways based on the observations in this study.

\vspace{1mm}
\noindent
\textbf{Potential recommendations for marketplace platforms:} Based on the insights of the study, here are few recommendations that may reduce such potential preferential treatments. 
First, the performance metrics should be evaluated in a uniform fashion for all sellers (irrespective of their special relationship with the marketplace) for a fair competition. This will not only provide more business opportunities to a wider set of sellers but also improve customers' trust on Amazon's algorithms. Second, there needs to be more widespread awareness of the different policies that Amazon adopts at different steps, since these policies ultimately affect the decision-making of customers. 
Although \#Ratings is not an indicator of quality, people seem to give more importance to it because it may be considered as a proxy for trust or reliable service (i.e., `so many people have bought from this seller'). 
This was seen in the survey that we conducted in Section~\ref{Sec: SurveyBuyBoxWithoutRateNum}.
Especially when we showed the number of ratings with the rectified metrics the percentage of first preference votes increases from 6.93\% to 28.53\%. 
So, complete removal of it might not be a good idea. However, Amazon can instead show how many units of that \textit{specific product} has been sold by the respective sellers. Although, it does not do away with the scale mismatch problem but it may balance the disparity in the scale to some extent.
Finally, Amazon should show and reveal its own performance metrics in USA, Germany and France and in all the countries where it can directly sell to customers. Much like third-party sellers, having a seller profile page may improve the transparency which can in fact improve customers' trust on Amazon or any e-commerce marketplace for that matter.

\color{black}
\vspace{1mm}
\noindent
\textbf{Takeaways for sellers on online marketplaces:} The current work bring to light some of the issues that sellers of different kinds face when they rely on marketplace platforms like Amazon. 
While most of the issues highlighted in this work are specific to the policies and algorithms deployed on the Amazon marketplaces, similar practices and choice architectures also exist in other e-commerce platforms such as Walmart and Alibaba. 
So, sellers need to be wary of these practices and policies deployed on e-commerce platforms. Another takeaway point is, irrespective of which delivery services they adopt (Amazon or third party), they need to make sure that their price and seller performance metrics need to be better for them to remain competitive in the eyes of customers. We hope the intervention of policymakers e.g., the lawsuit from FTC, the DMA regulations~\cite{EC2022DMA}, etc., into the matter may help the sellers' cause further. 

\vspace{1mm}
\noindent
\textbf{Takeaways for customers on online marketplaces:}
The choice architecture (deployed algorithms along with the policies) adopted by e-commerce marketplaces not only affect the opportunities of the sellers (by nudging customers toward \SpSeller{}) but also affect the customer satisfaction (e.g., by driving them away from an offer at a lower price by a less `accomplished' seller). Customers also need to be cautious about the different performance metrics that are shown on the marketplaces and their interpretation. Hopefully, going forward, researchers can play a role in raising such awareness among the general population about the nudges and algorithmic systems deployed on different online socio-technical platforms.

\vspace{1mm}
\noindent
\textbf{Takeaways for researchers and policymakers:}
Recently, the EU has agreed on a Digital Markets Act (DMA)~\cite{EC2022DMA} to regulate digital marketplaces such as Amazon for making it fair toward different stakeholders. 
This Act prohibits gatekeepers (such as Amazon) to treat services offered by itself more favorably than similar services of other third-party service providers.
However, in order to seamlessly deploy and check the adherence of these regulations, the gap between the laws and their technical operationalization needs to be thought through. 
This is where Computer Science / AI researchers who have the know-how of the technology and legal scholars who have the know-how of the regulations can come together and bridge such gaps.
We believe some of the observations and insights from this work can contribute to such discourse. Given the far reaching economic impact of such apparatus, they should be deployed with the interest of all the stakeholders in mind.
 
\color{black}

%% file: supplementary.tex
\appendix
\section{Data collection for \buybox{} competitions}\label{Sec: datacollectionBuyBox} 
One of the primary requirement for our study is to identify a set of products (preferably popular ones) for which there are multiple sellers competing for the \buybox{}. In addition, for each product the details of which seller(s) win the \buybox{} when the product page is visited a number of times is also required. Note that, visiting the same product pages multiple times is important to check whether 
the \buybox{} winners are 
rotated over time~\cite{chen2016empirical}.  
Given the popularity of product search on e-commerce platforms~\cite{sorokina2016amazon}, we proceeded to collect our dataset by navigating the product pages following the links from the search results. 
In particular, as a proof of concept, we collected data from all the Amazon marketplaces for a set of $100$ distinct most searched keywords (queries) on Amazon~\cite{Hardwick2021Top} across multiple popular categories.
The reason behind selecting such queries is twofold. First, since these queries are the most searched ones on Amazon, they enable us to land up on some of the most popular product pages visited by the customers. Second, through these queries we were able to come across multiple products where more than one sellers have offered the same product at different competitive prices. 

\vspace{1 mm}
\noindent
\textbf{Crawling methodology:} To collect the search results (and subsequently visit the product pages), we designed a crawler performing search operations on Amazon. The crawler was seeded with a  query, and the search results on the first SERP 
for the query were scraped. This process was repeated for all the queries. 
We used desktop browser automation for data collection. In particular, we used Gecko driver and Mozilla Firefox browser for data collection. 
We collected the data in different sessions from an Amazon account having prime membership. 
Therefore our choice is motivated to represent the behavior as observed by a majority of highly active Amazon customers~\cite{Dayton2020Amazon}.

We took some precautions during the data crawling process to ensure the meaningfulness of the analyses. The Amazon search algorithm is likely to tailor 
results based on several signals, including the geographical location, browsing history, etc. We also observed that search results and rankings vary over time. 
To minimize these variations, we collected search results in multiple temporal snapshots 
during the {\it same time frame}, from the {\it same IP address}, from the same user account, and from the same geographical location.

\section{Out of position \buybox{} win}
~\label{Sec: OOPBuybox}
\begin{figure}[t]
	\centering
	\begin{subfigure}{0.38\columnwidth}
		\includegraphics[width= \textwidth, height=4cm]{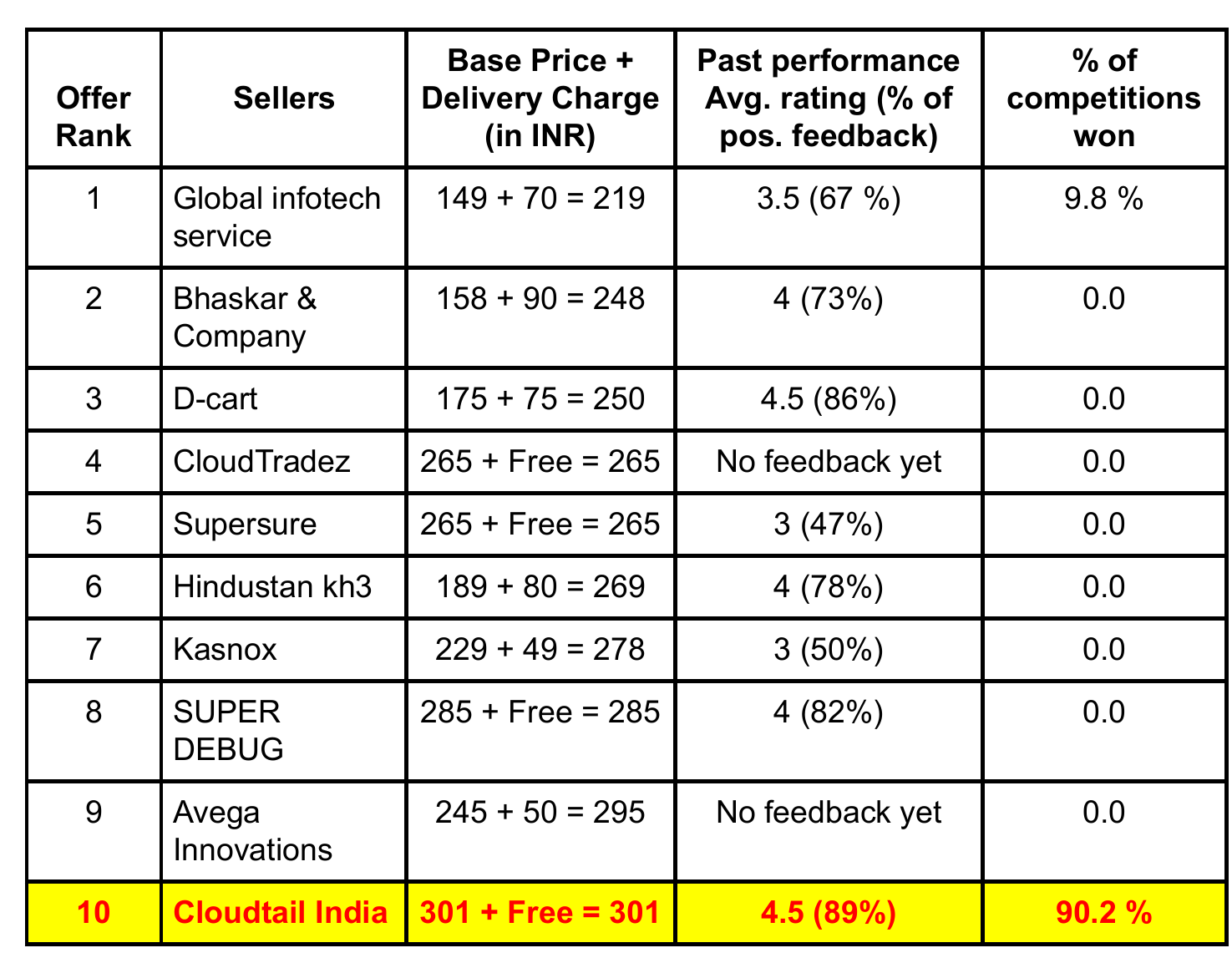}
		\vspace*{-3.5mm}
	\end{subfigure}%
	\vspace*{-3.5mm}
	\caption{{\bf An out of position \buybox{} win by an Amazon SM. The offer rank column shows the position in which the corresponding offers were shown as per price and delivery in the offer listing page. The seller marked in red (Cloudtail India -- an Amazon SM) was the winner of the \buybox{} in 90\% of the competitions for the product.}}
	\label{Fig: OutPositionInstance}
	\vspace{-4 mm}
	
\end{figure}

We show an example in Figure~\ref{Fig: OutPositionInstance}. The figure shows, for every offer (for the product), the total price (base price + delivery charges), performance metrics of the seller, and percentage of times the seller won the \buybox{} out of all the temporal snapshots we collected for this product. 
The `Offer rank' column denotes the order in which the offers were shown on the Amazon offer listing page at the time we collected the data (we show the top-10 offers only). 
Figure~\ref{Fig: OutPositionInstance} shows the set and ranking of offers which we observed in majority of the snapshots for this particular product. 
The seller highlighted in Figure~\ref{Fig: OutPositionInstance} (Cloudtail India, an Amazon SM) is the one that won the \buybox{} in more than 90\% of the snapshots when we encountered the corresponding products. 
Importantly, there were a number of sellers who offered the same product at lower prices and had comparable (or even better) performances as the SM. Despite this, a majority of the times the \buybox{} went to the SM.

\section{Data collection for seller feedback}~\label{Sec: DatacollectionSellerFeedback}
The other important dimension of the work is to investigate the effect of review striking-through on different sets of sellers on Amazon. To this end, we collected the seller feedback collected by 1000 active sellers on Amazon's Indian, American, German and French marketplaces. We extract the set of sellers from a list curated by Marketplace Pulse~\cite{MarketPlacePulsel2022Top}, where the sellers are sorted as per the number of feedback received in last 30 days. 

\vspace{1 mm}
\noindent
\textbf{Crawling methodology:} Much like the methodology adopted in the earlier data collection, here also we take help of browser automation. We used Gecko driver and Mozilla Firefox browser for data collection. Amazon arranges 5 seller feedback per page on its seller profile page. 
In Amazon's lingo, a feedback is a \textbf{$<$rating, review$>$} pair i.e., each review is associated with a rating out of 5 stars.
For each seller, we go to their Amazon seller profile pages and collect the feedback received by them in last 12 months or until when Amazon allows us to gather the data for the corresponding sellers. For some sellers, our crawler never went beyond a certain number of reviews even after multiple trials. Hence, we resorted to collect either the number of reviews mentioned for last 12 months or until when Amazon allows us to collect the data. The threshold was influenced by the fact that Amazon reports the \% of +ve feedback over past 12 months on offer listing pages.

%% file: Main.bbl

\begin{thebibliography}{57}


\ifx \showCODEN    \undefined \def \showCODEN     #1{\unskip}     \fi
\ifx \showDOI      \undefined \def \showDOI       #1{#1}\fi
\ifx \showISBNx    \undefined \def \showISBNx     #1{\unskip}     \fi
\ifx \showISBNxiii \undefined \def \showISBNxiii  #1{\unskip}     \fi
\ifx \showISSN     \undefined \def \showISSN      #1{\unskip}     \fi
\ifx \showLCCN     \undefined \def \showLCCN      #1{\unskip}     \fi
\ifx \shownote     \undefined \def \shownote      #1{#1}          \fi
\ifx \showarticletitle \undefined \def \showarticletitle #1{#1}   \fi
\ifx \showURL      \undefined \def \showURL       {\relax}        \fi
\providecommand\bibfield[2]{#2}
\providecommand\bibinfo[2]{#2}
\providecommand\natexlab[1]{#1}
\providecommand\showeprint[2][]{arXiv:#2}

\bibitem[Amazon(2021)]%
        {Amazon2021Becoming}
\bibfield{author}{\bibinfo{person}{Amazon}.} \bibinfo{year}{2021}\natexlab{}.
\newblock \bibinfo{title}{Becoming the Featured Offer}.
\newblock \bibinfo{howpublished}{https://amzn.to/3jVq2Vz}.
\newblock


\bibitem[Amazon(2023)]%
        {Amazon2023FBA}
\bibfield{author}{\bibinfo{person}{Amazon}.} \bibinfo{year}{2023}\natexlab{}.
\newblock \bibinfo{title}{Amazon FBA - Get the Prime advantage}.
\newblock \bibinfo{howpublished}{https://bit.ly/3RXqCSm}.
\newblock


\bibitem[Barocas and Selbst(2016)]%
        {barocas2016big}
\bibfield{author}{\bibinfo{person}{Solon Barocas} {and} \bibinfo{person}{Andrew~D Selbst}.} \bibinfo{year}{2016}\natexlab{}.
\newblock \showarticletitle{Big data's disparate impact}.
\newblock \bibinfo{journal}{\emph{Calif. L. Rev.}}  \bibinfo{volume}{104} (\bibinfo{year}{2016}), \bibinfo{pages}{671}.
\newblock


\bibitem[Burke(2017)]%
        {burke2017multisided}
\bibfield{author}{\bibinfo{person}{Robin Burke}.} \bibinfo{year}{2017}\natexlab{}.
\newblock \showarticletitle{Multisided fairness for recommendation}.
\newblock \bibinfo{journal}{\emph{arXiv preprint arXiv:1707.00093}} (\bibinfo{year}{2017}).
\newblock


\bibitem[Burke et~al\mbox{.}(2018)]%
        {burke2018balanced}
\bibfield{author}{\bibinfo{person}{Robin Burke}, \bibinfo{person}{Nasim Sonboli}, {and} \bibinfo{person}{Aldo Ordonez-Gauger}.} \bibinfo{year}{2018}\natexlab{}.
\newblock \showarticletitle{Balanced neighborhoods for multi-sided fairness in recommendation}. In \bibinfo{booktitle}{\emph{ACM FAccT}}.
\newblock


\bibitem[Calo(2013)]%
        {calo2013digital}
\bibfield{author}{\bibinfo{person}{Ryan Calo}.} \bibinfo{year}{2013}\natexlab{}.
\newblock \showarticletitle{Digital market manipulation}.
\newblock \bibinfo{journal}{\emph{Geo. Wash. L. Rev.}}  \bibinfo{volume}{82} (\bibinfo{year}{2013}), \bibinfo{pages}{995}.
\newblock


\bibitem[central(2022a)]%
        {SellerCentral2022Can}
\bibfield{author}{\bibinfo{person}{Amazon~Seller central}.} \bibinfo{year}{2022}\natexlab{a}.
\newblock \bibinfo{title}{Can Amazon remove buyer feedback?}
\newblock \bibinfo{howpublished}{https://bit.ly/3FLGMbu}.
\newblock


\bibitem[central(2022b)]%
        {SellerCentral2022BuyShip}
\bibfield{author}{\bibinfo{person}{Amazon~Seller central}.} \bibinfo{year}{2022}\natexlab{b}.
\newblock \bibinfo{title}{Use Buy Shipping services}.
\newblock \bibinfo{howpublished}{https://bit.ly/41Rlnbx}.
\newblock


\bibitem[central(2022c)]%
        {SellerCentral2022Buyer}
\bibfield{author}{\bibinfo{person}{Amazon~Seller central}.} \bibinfo{year}{2022}\natexlab{c}.
\newblock \bibinfo{title}{What happens when I receive a negative feedback on an AFN order?}
\newblock \bibinfo{howpublished}{https://bit.ly/4aOvbqD}.
\newblock


\bibitem[Chen et~al\mbox{.}(2016)]%
        {chen2016empirical}
\bibfield{author}{\bibinfo{person}{Le Chen}, \bibinfo{person}{Alan Mislove}, {and} \bibinfo{person}{Christo Wilson}.} \bibinfo{year}{2016}\natexlab{}.
\newblock \showarticletitle{An empirical analysis of algorithmic pricing on amazon marketplace}. In \bibinfo{booktitle}{\emph{WWW}}.
\newblock


\bibitem[Commision(2022)]%
        {EC2022DMA}
\bibfield{author}{\bibinfo{person}{European Commision}.} \bibinfo{year}{2022}\natexlab{}.
\newblock \bibinfo{title}{Digital Markets Act (DMA)}.
\newblock \bibinfo{howpublished}{https://eur-lex.europa.eu/eli/reg/2022/1925}.
\newblock


\bibitem[Dalal and Verma(2016)]%
        {Dalal2016Flipkart}
\bibfield{author}{\bibinfo{person}{Mihir Dalal} {and} \bibinfo{person}{Shrutika Verma}.} \bibinfo{year}{2016}\natexlab{}.
\newblock \bibinfo{title}{Flipkart looks to shift most of its sales to a few sellers}.
\newblock \bibinfo{howpublished}{https://bit.ly/3O4BQmL}.
\newblock


\bibitem[Dash et~al\mbox{.}(2021)]%
        {dash2021when}
\bibfield{author}{\bibinfo{person}{Abhisek Dash}, \bibinfo{person}{Abhijnan Chakraborty}, \bibinfo{person}{Saptarshi Ghosh}, \bibinfo{person}{Animesh Mukherjee}, {and} \bibinfo{person}{Krishna~P. Gummadi}.} \bibinfo{year}{2021}\natexlab{}.
\newblock \showarticletitle{When the Umpire is also a Player: Bias in Private Label Product Recommendations on E-commerce Marketplaces}. In \bibinfo{booktitle}{\emph{ACM FAccT}}.
\newblock


\bibitem[Dash et~al\mbox{.}(2022)]%
        {dash2022alexa}
\bibfield{author}{\bibinfo{person}{Abhisek Dash}, \bibinfo{person}{Abhijnan Chakraborty}, \bibinfo{person}{Saptarshi Ghosh}, \bibinfo{person}{Animesh Mukherjee}, {and} \bibinfo{person}{Krishna~P Gummadi}.} \bibinfo{year}{2022}\natexlab{}.
\newblock \showarticletitle{Alexa, in you, I trust! Fairness and Interpretability Issues in E-commerce Search through Smart Speakers}. In \bibinfo{booktitle}{\emph{ACM Web Conference}}.
\newblock


\bibitem[Dayton(2020)]%
        {Dayton2020Amazon}
\bibfield{author}{\bibinfo{person}{Emily Dayton}.} \bibinfo{year}{2020}\natexlab{}.
\newblock \bibinfo{title}{Amazon Statistics You Should Know: Opportunities to Make the Most of America's Top Online Marketplace}.
\newblock \bibinfo{howpublished}{https://bit.ly/2WWEFP8}.
\newblock


\bibitem[Edgerton and Nylen(2023)]%
        {Edgerton2023Lina}
\bibfield{author}{\bibinfo{person}{Anna Edgerton} {and} \bibinfo{person}{Leah Nylen}.} \bibinfo{year}{2023}\natexlab{}.
\newblock \bibinfo{title}{Lina Khan Is Coming for Amazon, Armed With an FTC Antitrust Suit}.
\newblock \bibinfo{howpublished}{https://www.bloomberg.com/news/articles/2023-06-29/amazon-major-ftc-antitrust-case-expected-in-coming-weeks}.
\newblock


\bibitem[FTC(2023)]%
        {FTC2023AmazonSue}
\bibfield{author}{\bibinfo{person}{USA FTC}.} \bibinfo{year}{2023}\natexlab{}.
\newblock \bibinfo{title}{FTC Sues Amazon for Illegally Maintaining Monopoly Power}.
\newblock \bibinfo{howpublished}{https://www.ftc.gov/news-events/news/press-releases/2023/09/ftc-sues-amazon-illegally-maintaining-monopoly-power}.
\newblock


\bibitem[Geyik et~al\mbox{.}(2019)]%
        {geyik2019fairness}
\bibfield{author}{\bibinfo{person}{Sahin~Cem Geyik}, \bibinfo{person}{Stuart Ambler}, {and} \bibinfo{person}{Krishnaram Kenthapadi}.} \bibinfo{year}{2019}\natexlab{}.
\newblock \bibinfo{title}{Fairness-Aware Ranking in Search \& Recommendation Systems with Application to LinkedIn Talent Search}.
\newblock
\newblock


\bibitem[Gupta et~al\mbox{.}(2023)]%
        {gupta2023towards}
\bibfield{author}{\bibinfo{person}{Anjali Gupta}, \bibinfo{person}{Shreyans~J Nagori}, \bibinfo{person}{Abhijnan Chakraborty}, \bibinfo{person}{Rohit Vaish}, \bibinfo{person}{Sayan Ranu}, \bibinfo{person}{Prajit~Prashant Nadkarni}, \bibinfo{person}{Narendra~Varma Dasararaju}, {and} \bibinfo{person}{Muthusamy Chelliah}.} \bibinfo{year}{2023}\natexlab{}.
\newblock \showarticletitle{Towards Fair Allocation in Social Commerce Platforms}. In \bibinfo{booktitle}{\emph{Proceedings of the ACM Web Conference 2023}}. \bibinfo{pages}{3744--3754}.
\newblock


\bibitem[Hanson and Kysar(1999)]%
        {hanson1999taking}
\bibfield{author}{\bibinfo{person}{Jon~D Hanson} {and} \bibinfo{person}{Douglas~A Kysar}.} \bibinfo{year}{1999}\natexlab{}.
\newblock \showarticletitle{Taking behavioralism seriously: The problem of market manipulation}.
\newblock \bibinfo{journal}{\emph{HeinOnline NYUL rev.}}  \bibinfo{volume}{74} (\bibinfo{year}{1999}).
\newblock


\bibitem[Hardwick(2021)]%
        {Hardwick2021Top}
\bibfield{author}{\bibinfo{person}{Joshua Hardwick}.} \bibinfo{year}{2021}\natexlab{}.
\newblock \bibinfo{title}{Top Amazon Searches (2021)}.
\newblock \bibinfo{howpublished}{https://ahrefs.com/blog/top-amazon-searches/}.
\newblock


\bibitem[He et~al\mbox{.}(2022)]%
        {he2022market}
\bibfield{author}{\bibinfo{person}{Sherry He}, \bibinfo{person}{Brett Hollenbeck}, {and} \bibinfo{person}{Davide Proserpio}.} \bibinfo{year}{2022}\natexlab{}.
\newblock \showarticletitle{The market for fake reviews}.
\newblock \bibinfo{journal}{\emph{INFORMS Marketing Science}} \bibinfo{volume}{41}, \bibinfo{number}{5} (\bibinfo{year}{2022}).
\newblock


\bibitem[House Committee on~the Judiciary(2020)]%
        {House2020Judiciary}
\bibfield{author}{\bibinfo{person}{USA House Committee on~the Judiciary}.} \bibinfo{year}{2020}\natexlab{}.
\newblock \bibinfo{title}{Judiciary Antitrust Subcommittee Investigation Reveals Digital Economy Highly Concentrated, Impacted By Monopoly Power}.
\newblock \bibinfo{howpublished}{https://bit.ly/3yQmlo3}.
\newblock


\bibitem[Hussein et~al\mbox{.}(2020)]%
        {hussein2020measuring}
\bibfield{author}{\bibinfo{person}{Eslam Hussein}, \bibinfo{person}{Prerna Juneja}, {and} \bibinfo{person}{Tanushree Mitra}.} \bibinfo{year}{2020}\natexlab{}.
\newblock \showarticletitle{Measuring misinformation in video search platforms: An audit study on YouTube}.
\newblock \bibinfo{journal}{\emph{PACM HCI (CSCW)}} (\bibinfo{year}{2020}).
\newblock


\bibitem[Johnson et~al\mbox{.}(2012)]%
        {johnson2012beyond}
\bibfield{author}{\bibinfo{person}{Eric~J Johnson}, \bibinfo{person}{Suzanne~B Shu}, \bibinfo{person}{Benedict~GC Dellaert}, \bibinfo{person}{Craig Fox}, \bibinfo{person}{Daniel~G Goldstein}, \bibinfo{person}{Gerald H{\"a}ubl}, \bibinfo{person}{Richard~P Larrick}, \bibinfo{person}{John~W Payne}, \bibinfo{person}{Ellen Peters}, \bibinfo{person}{David Schkade}, {et~al\mbox{.}}} \bibinfo{year}{2012}\natexlab{}.
\newblock \showarticletitle{Beyond nudges: Tools of a choice architecture}.
\newblock \bibinfo{journal}{\emph{Springer Marketing Letters}} (\bibinfo{year}{2012}).
\newblock


\bibitem[Juneja and Mitra(2021)]%
        {juneja2021auditing}
\bibfield{author}{\bibinfo{person}{Prerna Juneja} {and} \bibinfo{person}{Tanushree Mitra}.} \bibinfo{year}{2021}\natexlab{}.
\newblock \showarticletitle{Auditing e-commerce platforms for algorithmically curated vaccine misinformation}. In \bibinfo{booktitle}{\emph{ACM SIGCHI}}. \bibinfo{pages}{1--27}.
\newblock


\bibitem[Kahneman et~al\mbox{.}(1982)]%
        {kahneman1982judgment}
\bibfield{author}{\bibinfo{person}{Daniel Kahneman}, \bibinfo{person}{Paul Slovic}, {and} \bibinfo{person}{Amos Tversky}.} \bibinfo{year}{1982}\natexlab{}.
\newblock \bibinfo{booktitle}{\emph{Judgment under uncertainty: Heuristics and biases}}.
\newblock \bibinfo{publisher}{Cambridge university press}.
\newblock


\bibitem[Kalra(2021a)]%
        {Kalra2021Amazon}
\bibfield{author}{\bibinfo{person}{Aditya Kalra}.} \bibinfo{year}{2021}\natexlab{a}.
\newblock \bibinfo{title}{Amazon documents reveal company's secret strategy to dodge India's regulators}.
\newblock \bibinfo{howpublished}{https://reut.rs/2Yt3GlL}.
\newblock


\bibitem[Kalra(2021b)]%
        {Kalra2021India2}
\bibfield{author}{\bibinfo{person}{Aditya Kalra}.} \bibinfo{year}{2021}\natexlab{b}.
\newblock \bibinfo{title}{India's draft e-commerce policy calls for equal treatment of sellers}.
\newblock \bibinfo{howpublished}{https://reut.rs/3yNV7P6}.
\newblock


\bibitem[Kaziukenas(2019)]%
        {Kaziukenas2019Amazon}
\bibfield{author}{\bibinfo{person}{J.~J Kaziukenas}.} \bibinfo{year}{2019}\natexlab{}.
\newblock \bibinfo{title}{Amazon Has Three Million Active Sellers}.
\newblock \bibinfo{howpublished}{https://bit.ly/38JhfiL}.
\newblock


\bibitem[Khan(2016)]%
        {khan2016amazon}
\bibfield{author}{\bibinfo{person}{Lina~M Khan}.} \bibinfo{year}{2016}\natexlab{}.
\newblock \showarticletitle{Amazon's antitrust paradox}.
\newblock \bibinfo{journal}{\emph{Yale LJ}} (\bibinfo{year}{2016}).
\newblock


\bibitem[Kulshrestha et~al\mbox{.}(2017)]%
        {kulshrestha2017quantifying}
\bibfield{author}{\bibinfo{person}{Juhi Kulshrestha}, \bibinfo{person}{Motahhare Eslami}, \bibinfo{person}{Johnnatan Messias}, \bibinfo{person}{Muhammad~Bilal Zafar}, \bibinfo{person}{Saptarshi Ghosh}, \bibinfo{person}{Krishna~P Gummadi}, {and} \bibinfo{person}{Karrie Karahalios}.} \bibinfo{year}{2017}\natexlab{}.
\newblock \showarticletitle{Quantifying search bias: Investigating sources of bias for political searches in social media}. In \bibinfo{booktitle}{\emph{ACM CSCW}}.
\newblock


\bibitem[Lanxner(2021)]%
        {Lanxner2021Amazon}
\bibfield{author}{\bibinfo{person}{Eyal Lanxner}.} \bibinfo{year}{2021}\natexlab{}.
\newblock \bibinfo{title}{The Amazon Buy Box: How It Works for Sellers, and Why It's So Important}.
\newblock \bibinfo{howpublished}{https://bit.ly/3h5a21a}.
\newblock


\bibitem[Luo et~al\mbox{.}(2018)]%
        {luo2018online}
\bibfield{author}{\bibinfo{person}{Sheng Luo}, \bibinfo{person}{Bin Gu}, \bibinfo{person}{Xingbiao Wang}, {and} \bibinfo{person}{Zhaoquan Zhou}.} \bibinfo{year}{2018}\natexlab{}.
\newblock \showarticletitle{Online compulsive buying behavior: The mediating role of self-control and negative emotions}. In \bibinfo{booktitle}{\emph{ICEBI}}. \bibinfo{pages}{65--69}.
\newblock


\bibitem[Malik(2021)]%
        {Malik2021Alpha}
\bibfield{author}{\bibinfo{person}{Bismah Malik}.} \bibinfo{year}{2021}\natexlab{}.
\newblock \bibinfo{title}{How alpha sellers have put Amazon, Flipkart in dock?}
\newblock \bibinfo{howpublished}{https://bit.ly/3U0xiBB}.
\newblock


\bibitem[MarketllacePulse(2022)]%
        {MarketPlacePulsel2022Top}
\bibfield{author}{\bibinfo{person}{MarketllacePulse}.} \bibinfo{year}{2022}\natexlab{}.
\newblock \bibinfo{title}{Top Amazon.in Marketplace Sellers}.
\newblock \bibinfo{howpublished}{https://www.marketplacepulse.com/amazon/top-amazon-india-sellers}.
\newblock


\bibitem[Mathew et~al\mbox{.}(2020)]%
        {mathew2020hate}
\bibfield{author}{\bibinfo{person}{Binny Mathew}, \bibinfo{person}{Anurag Illendula}, \bibinfo{person}{Punyajoy Saha}, \bibinfo{person}{Soumya Sarkar}, \bibinfo{person}{Pawan Goyal}, {and} \bibinfo{person}{Animesh Mukherjee}.} \bibinfo{year}{2020}\natexlab{}.
\newblock \showarticletitle{Hate begets hate: A temporal study of hate speech}.
\newblock \bibinfo{journal}{\emph{PACM HCI (CSCW)}} (\bibinfo{year}{2020}).
\newblock


\bibitem[Mathur et~al\mbox{.}(2019)]%
        {mathur2019dark}
\bibfield{author}{\bibinfo{person}{Arunesh Mathur}, \bibinfo{person}{Gunes Acar}, \bibinfo{person}{Michael~J Friedman}, \bibinfo{person}{Eli Lucherini}, \bibinfo{person}{Jonathan Mayer}, \bibinfo{person}{Marshini Chetty}, {and} \bibinfo{person}{Arvind Narayanan}.} \bibinfo{year}{2019}\natexlab{}.
\newblock \showarticletitle{Dark patterns at scale: Findings from a crawl of 11K shopping websites}.
\newblock \bibinfo{journal}{\emph{PACM HCI}} \bibinfo{volume}{3}, \bibinfo{number}{CSCW} (\bibinfo{year}{2019}).
\newblock


\bibitem[Mehrotra et~al\mbox{.}(2018)]%
        {mehrotra2018towards}
\bibfield{author}{\bibinfo{person}{Rishabh Mehrotra}, \bibinfo{person}{James McInerney}, \bibinfo{person}{Hugues Bouchard}, \bibinfo{person}{Mounia Lalmas}, {and} \bibinfo{person}{Fernando Diaz}.} \bibinfo{year}{2018}\natexlab{}.
\newblock \showarticletitle{Towards a fair marketplace: Counterfactual evaluation of the trade-off between relevance, fairness \& satisfaction in recommendation systems}. In \bibinfo{booktitle}{\emph{ACM CIKM}}.
\newblock


\bibitem[Mehrotra et~al\mbox{.}(2020)]%
        {mehrotra2020bandit}
\bibfield{author}{\bibinfo{person}{Rishabh Mehrotra}, \bibinfo{person}{Niannan Xue}, {and} \bibinfo{person}{Mounia Lalmas}.} \bibinfo{year}{2020}\natexlab{}.
\newblock \showarticletitle{Bandit based optimization of multiple objectives on a music streaming platform}. In \bibinfo{booktitle}{\emph{ACM SIGKDD}}.
\newblock


\bibitem[Moser et~al\mbox{.}(2019)]%
        {moser2019impulse}
\bibfield{author}{\bibinfo{person}{Carol Moser}, \bibinfo{person}{Sarita~Y Schoenebeck}, {and} \bibinfo{person}{Paul Resnick}.} \bibinfo{year}{2019}\natexlab{}.
\newblock \showarticletitle{Impulse buying: Design practices and consumer needs}. In \bibinfo{booktitle}{\emph{ACM SIGCHI}}.
\newblock


\bibitem[Mota et~al\mbox{.}(2020)]%
        {mota2020desiderata}
\bibfield{author}{\bibinfo{person}{Nuno Mota}, \bibinfo{person}{Abhijnan Chakraborty}, \bibinfo{person}{Asia~J Biega}, \bibinfo{person}{Krishna~P Gummadi}, {and} \bibinfo{person}{Hoda Heidari}.} \bibinfo{year}{2020}\natexlab{}.
\newblock \showarticletitle{On the desiderata for online altruism: Nudging for equitable donations}.
\newblock \bibinfo{journal}{\emph{ACM PHCI}} \bibinfo{volume}{4}, \bibinfo{number}{CSCW2} (\bibinfo{year}{2020}).
\newblock


\bibitem[Patro et~al\mbox{.}(2020a)]%
        {patro2020fairrec}
\bibfield{author}{\bibinfo{person}{Gourab~K Patro}, \bibinfo{person}{Arpita Biswas}, \bibinfo{person}{Niloy Ganguly}, \bibinfo{person}{Krishna~P Gummadi}, {and} \bibinfo{person}{Abhijnan Chakraborty}.} \bibinfo{year}{2020}\natexlab{a}.
\newblock \showarticletitle{FairRec: Two-Sided Fairness for Personalized Recommendations in Two-Sided Platforms}. In \bibinfo{booktitle}{\emph{WWW}}.
\newblock


\bibitem[Patro et~al\mbox{.}(2020b)]%
        {patro2020incremental}
\bibfield{author}{\bibinfo{person}{Gourab~K Patro}, \bibinfo{person}{Abhijnan Chakraborty}, \bibinfo{person}{Niloy Ganguly}, {and} \bibinfo{person}{Krishna~P Gummadi}.} \bibinfo{year}{2020}\natexlab{b}.
\newblock \showarticletitle{Incremental Fairness in Two-Sided Market Platforms: On Smoothly Updating Recommendations}.
\newblock \bibinfo{journal}{\emph{AAAI}} (\bibinfo{year}{2020}).
\newblock


\bibitem[PIB(2018)]%
        {Govt2018FDI}
\bibfield{author}{\bibinfo{person}{PIB}.} \bibinfo{year}{2018}\natexlab{}.
\newblock \bibinfo{title}{Review of policy on Foreign Direct Investment (FDI) in e-commerce}.
\newblock \bibinfo{howpublished}{https://bit.ly/3BP8vnO}.
\newblock


\bibitem[Prolific(2023)]%
        {Prolific2023Payment}
\bibfield{author}{\bibinfo{person}{Prolific}.} \bibinfo{year}{2023}\natexlab{}.
\newblock \bibinfo{title}{How much should you pay research participants?}
\newblock \bibinfo{howpublished}{https://bit.ly/47ALRiv}.
\newblock


\bibitem[Release(2020)]%
        {EU2020Antitrust}
\bibfield{author}{\bibinfo{person}{EU~Press Release}.} \bibinfo{year}{2020}\natexlab{}.
\newblock \bibinfo{title}{Antitrust: Commission sends Statement of Objections to Amazon for the use of non-public independent seller data and opens second investigation into its e-commerce business practices}.
\newblock \bibinfo{howpublished}{https://bit.ly/3h8jAIR}.
\newblock


\bibitem[Response(2019)]%
        {Amazon2019Online}
\bibfield{author}{\bibinfo{person}{Amazon's Response}.} \bibinfo{year}{2019}\natexlab{}.
\newblock \bibinfo{title}{Online Platforms and Market Power, Part 2: Innovation and Entrepreneurship}.
\newblock \bibinfo{howpublished}{https://bit.ly/2VnSS7q}.
\newblock


\bibitem[Response(2020)]%
        {Amazon2020Questions}
\bibfield{author}{\bibinfo{person}{Amazon's Response}.} \bibinfo{year}{2020}\natexlab{}.
\newblock \bibinfo{title}{Hearing of the Subcommittee on Antitrust, Commercial, and Administrative Law, Committee on the Judiciary}.
\newblock \bibinfo{howpublished}{https://bit.ly/3n9Um0k}.
\newblock


\bibitem[Saha et~al\mbox{.}(2021)]%
        {saha2021short}
\bibfield{author}{\bibinfo{person}{Punyajoy Saha}, \bibinfo{person}{Binny Mathew}, \bibinfo{person}{Kiran Garimella}, {and} \bibinfo{person}{Animesh Mukherjee}.} \bibinfo{year}{2021}\natexlab{}.
\newblock \showarticletitle{“Short is the Road that Leads from Fear to Hate”: Fear Speech in Indian WhatsApp Groups}. In \bibinfo{booktitle}{\emph{Proceedings of the Web conference 2021}}.
\newblock


\bibitem[Schmitt(2023)]%
        {Schmitt2023Amazon}
\bibfield{author}{\bibinfo{person}{Nicola Schmitt}.} \bibinfo{year}{2023}\natexlab{}.
\newblock \bibinfo{title}{With Amazon Around the World – The Amazon Marketplaces Worldwide in Focus}.
\newblock \bibinfo{howpublished}{https://www.blankspace.eu/blog-posts-en/amazon-marketplaces-worldwide}.
\newblock


\bibitem[Schneider et~al\mbox{.}(2018)]%
        {schneider2018digital}
\bibfield{author}{\bibinfo{person}{Christoph Schneider}, \bibinfo{person}{Markus Weinmann}, {and} \bibinfo{person}{Jan vom Brocke}.} \bibinfo{year}{2018}\natexlab{}.
\newblock \showarticletitle{Digital nudging: guiding online user choices through interface design}.
\newblock \bibinfo{journal}{\emph{ACM CACM}} (\bibinfo{year}{2018}).
\newblock


\bibitem[Sorokina and Cantu-Paz(2016)]%
        {sorokina2016amazon}
\bibfield{author}{\bibinfo{person}{Daria Sorokina} {and} \bibinfo{person}{Erick Cantu-Paz}.} \bibinfo{year}{2016}\natexlab{}.
\newblock \showarticletitle{Amazon search: The joy of ranking products}. In \bibinfo{booktitle}{\emph{ACM SIGIR}}.
\newblock


\bibitem[Thaler and Sunstein(2008)]%
        {thaler2008nudge}
\bibfield{author}{\bibinfo{person}{RH Thaler} {and} \bibinfo{person}{CR Sunstein}.} \bibinfo{year}{2008}\natexlab{}.
\newblock \bibinfo{booktitle}{\emph{Nudge: improving decisions about health, wealth and happiness Penguin}}.
\newblock \bibinfo{publisher}{Penguin Books, New York}.
\newblock


\bibitem[Tversky and Kahneman(1981)]%
        {tversky1981framing}
\bibfield{author}{\bibinfo{person}{Amos Tversky} {and} \bibinfo{person}{Daniel Kahneman}.} \bibinfo{year}{1981}\natexlab{}.
\newblock \showarticletitle{The framing of decisions and the psychology of choice}.
\newblock \bibinfo{journal}{\emph{science}} \bibinfo{volume}{211}, \bibinfo{number}{4481} (\bibinfo{year}{1981}).
\newblock


\bibitem[University(2023)]%
        {SellerCentral2022Feedback}
\bibfield{author}{\bibinfo{person}{Amazon~Seller University}.} \bibinfo{year}{2023}\natexlab{}.
\newblock \bibinfo{title}{About Feedback Manager}.
\newblock \bibinfo{howpublished}{https://sell.amazon.in/sell-online/fulfillment-by-amazon}.
\newblock


\bibitem[Yin and Jeffries(2021)]%
        {Yin2021Search}
\bibfield{author}{\bibinfo{person}{Leon Yin} {and} \bibinfo{person}{Adrianne Jeffries}.} \bibinfo{year}{2021}\natexlab{}.
\newblock \bibinfo{title}{How We Analyzed Amazon's Treatment of Its ``Brands'' in Search Results}.
\newblock \bibinfo{howpublished}{https://bit.ly/420SiKG}.
\newblock


\end{thebibliography}
